\documentclass[11pt,letterpaper]{article}
\pdfoutput=1
\usepackage{jcappub}
\usepackage{subcaption}
\usepackage{bbm,bm}
\usepackage{mathrsfs}
\usepackage{slashed}
\usepackage{caption}
\usepackage{epstopdf}
\usepackage[normalem]{ulem}
\usepackage[bottom]{footmisc}
\usepackage{subcaption}
\usepackage{bbold}
\usepackage{titlesec}
\usepackage{threeparttable}
\usepackage{booktabs}
\usepackage{changepage}
\usepackage[utf8]{inputenc}
\usepackage{soul}
\usepackage{accents}

\usepackage{mhchem}
\usepackage{grffile}

\usepackage{graphicx}  
\usepackage{dcolumn}   
\usepackage{bm}        
\usepackage{amssymb}   
\usepackage{setspace}
\usepackage{amsmath, amssymb, setspace}
\usepackage{array}
\usepackage{booktabs}
\usepackage{caption}
\usepackage{indentfirst}
\usepackage{float}
\usepackage{lmodern}
\usepackage{multirow}
\usepackage{soul}
\usepackage[normalem]{ulem}

\usepackage{braket}
\usepackage{comment}

\usepackage[draft]{pgf}

\usepackage{tikz}
\usepackage{xcolor}

\newcommand{\DoBox}[1]{\begin{center}
\color{red}\fbox{
\begin{minipage}{0.9\textwidth}

\end{minipage}}
\end{center}}

\newcommand{\ie}{{i.e.}}

\newcommand{\eg}{{e.g.}}

\newcommand{\fig}{Fig.}

\newcommand{\Refe}{Ref.}
\newcommand{\Refes}{Refs.}

\newcommand{\figu}[1]{\fig~\ref{fig:#1}}

\definecolor{forestgreen}{rgb}{0,0.6,0}

\newlength{\myimageoversize}
\newsavebox{\myimage}

\titleformat{\subsubsection}
 {\normalfont\fontsize{12}{17}\itshape}{\thesubsubsection}{1em}{}

\title{\huge{The next galactic supernova can uncover mass and couplings of particles decaying to neutrinos}}
\author[a]{Bernanda Telalovic,}
\author[a]{Damiano F.G. Fiorillo,}
\author[a,b]{Pablo Martínez-Miravé,}
\author[c,d]{Edoardo Vitagliano,}
\author[a]{Mauricio Bustamante}

\affiliation[a]{Niels Bohr International Academy, Niels Bohr Institute, \\ University of Copenhagen, Blegdamsvej 17, 2100 Copenhagen, Denmark
}
\affiliation[b]{DARK, Niels Bohr Institute, \\ University of Copenhagen, Jagtvej 128, 2200 Copenhagen, Denmark
}

\affiliation[c]{Dipartimento di Fisica e Astronomia, \\ Università degli Studi di Padova, Via Marzolo 8, 35131 Padova, Italy}
\affiliation[d]{Istituto Nazionale di Fisica Nucleare (INFN),\\ Sezione di Padova, Via Marzolo 8, 35131 Padova, Italy}

\emailAdd{bernanda.telalovic@nbi.ku.dk}
\emailAdd{damiano.fiorillo@nbi.ku.dk}
\emailAdd{pablo.mirave@nbi.ku.dk}
\emailAdd{edoardo.vitagliano@unipd.it}
\emailAdd{mbustamante@nbi.ku.dk}
\begin{document}

\abstract{Many particles predicted by extensions of the Standard Model feature interactions with neutrinos, e.g., Majoron-like bosons $\phi$. If the mass of $\phi$ is larger than about $10\,\rm keV$, they can be produced abundantly in the core of the next galactic core-collapse supernova through neutrino coalescence, and leave it with energies of around 100~MeV. Their subsequent decay to high-energy neutrinos and anti-neutrinos provides a distinctive signature at Earth. Ongoing and planned neutrino and dark matter experiments allow us to reconstruct the energy, flavor, and time of arrival of these high-energy neutrinos. For the first time, we show that these measurements can help pinpointing the mass of $\phi$ and its couplings to neutrinos of different flavor. Our results can be generalized in a straightforward manner to other hypothetical feebly interacting particles, like novel gauge bosons or heavy neutral leptons, that decay into neutrinos.
}

\maketitle

\section{Introduction}
\label{sec:introduction}

Because neutrinos interact weakly with other particles of the Standard Model (SM), their properties are the least understood, making them compelling candidates to act as portals to new physics~\cite{Mohapatra:2005wg, Cirigliano:2013lpa, Berryman:2022hds}.
Many proposed extensions of the SM posit new particles that couple to neutrinos, including new neutral leptons~\cite{Asaka:2005pn, Gorbunov:2007ak}, bosons associated with novel gauge symmetries~\cite{Gell-Mann:1976xin, Witten:1979nr, Mohapatra:1979ia, Chikashige:1980ht}, and Goldstone and pseudo-Goldstone bosons associated with the spontaneous breaking of global symmetries~\cite{Chikashige:1980ui, Gelmini:1980re, Schechter:1981cv, Georgi:1981pg}. The search for physics beyond the SM (BSM) through neutrinos spans astrophysical, cosmological, and laboratory observations; for a necessarily limited list, see \Refes~\cite{Zeldovich:1980st, Gronau:1984ct, Kim:1986ax, Carlson:1986cu, Raffelt:1996wa, DELPHI:1996qcc, Johnson:1997cj, NuTeV:1999kej, Dolgov:2000jw, Kachelriess:2000qc, L3:2001xsz, Farzan:2002wx, EXO-200:2014vam, Hardy:2016kme, CMS:2018iaf, Bondarenko:2018ptm, Kling:2018wct, Magill:2018jla, Ballett:2019bgd, ATLAS:2019kpx, DeGouvea:2019wpf, Mastrototaro:2019vug, Escudero:2019gvw, Kelly:2020aks, Brdar:2020quo, Arguelles:2021dqn, Kharusi:2021jez, CUPID-0:2022yws, Abdullahi:2022jlv, Fiorillo:2022cdq, Fiorillo:2023ytr, Carenza:2023old, Brdar:2023tmi, Fiorillo:2023cas, Sandner:2023ptm, Chauhan:2023sci}.

Together with the early universe, the hot, dense cores of collapsing stars are the only places where neutrinos can reach thermal equilibrium. At such large temperatures and densities, supernovae (SNe) might be powerful factories of new particles that couple to neutrinos. These include Majoron-like bosons, $\phi$, hereafter simply called Majorons, whose interaction with neutrinos is of the form (see, \eg, \Refe~\cite{Farzan:2002wx})
\begin{equation}\label{eq:Lagrangian}
    \mathcal{L}_{\text{int}}=-\sum_{\alpha}\frac{g_{\alpha}}{2}\,\psi_\alpha^T\sigma_2\psi_\alpha \phi+\textrm{h.c.} \;,
\end{equation}
where $\psi_\alpha$ is the two-component Majorana neutrino field with flavor $\alpha=e,\mu,\tau$, and $g_{\alpha}$ is the coupling, which must be a real number. For simplicity, we restrict ourselves to the case of flavor-diagonal couplings. The above coupling entails lepton number violation.

If Majorons have a mass $m_\phi\gtrsim 10\,\rm keV$,\footnote{Throughout this work, we use natural units such that $\hbar=c=1$.} they could be produced abundantly in the cores of SNe through neutrino coalescence, $\nu\nu\to \phi$. These Majorons can leave the core with energies of hundreds of MeV, take energy away from the core, and shorten the duration of the standard neutrino signal expected at neutrino detectors. As a result, one can infer neutrino cooling bounds on $\phi$---akin to those on the QCD axion~\cite{Chang:2018rso, Carenza:2019pxu, Carenza:2020cis, Caputo:2024oqc}, axion-like particles~\cite{Bollig:2020xdr, Croon:2020lrf, Caputo:2021rux, Caputo:2022rca, Caputo:2022mah, Ferreira:2022xlw}, dark photons~\cite{DeRocco:2019njg, Shin:2022ulh}, or millicharged particles~\cite{Davidson:2000hf, Fiorillo:2024upk}---by looking at the time distribution of the SN 1987A neutrinos detected by Kamiokande~II~\cite{Kamiokande-II:1987idp, Hirata:1988ad, Hirata:1991td, Koshiba:1992yb, Oyama:2021oqp}, the Irvine–Michigan–Brookhaven~\cite{Bionta:1987qt, 1987svoboda, IMB:1988suc}, and the Baksan Underground Scintillation Telescope (BUST) \cite{Alekseev:1987ej, Alekseev:1988gp}. However, it was pointed out in \Refe~\cite{Fiorillo:2022cdq} that $\phi$ could have decayed back to neutrinos away from the cooling proto-neutron star, producing a flux of neutrinos with unexpectedly high energies compared to the standard tens-of-MeV neutrinos emitted from the neutrinosphere. Since such flux was not observed from SN~1987A, one can draw constraints which are always stronger than the cooling bounds. (See also \Refe~\cite{Akita:2023iwq} for an application to a list of models.)
\begin{figure}
    \centering
    \includegraphics[width=0.85\textwidth]{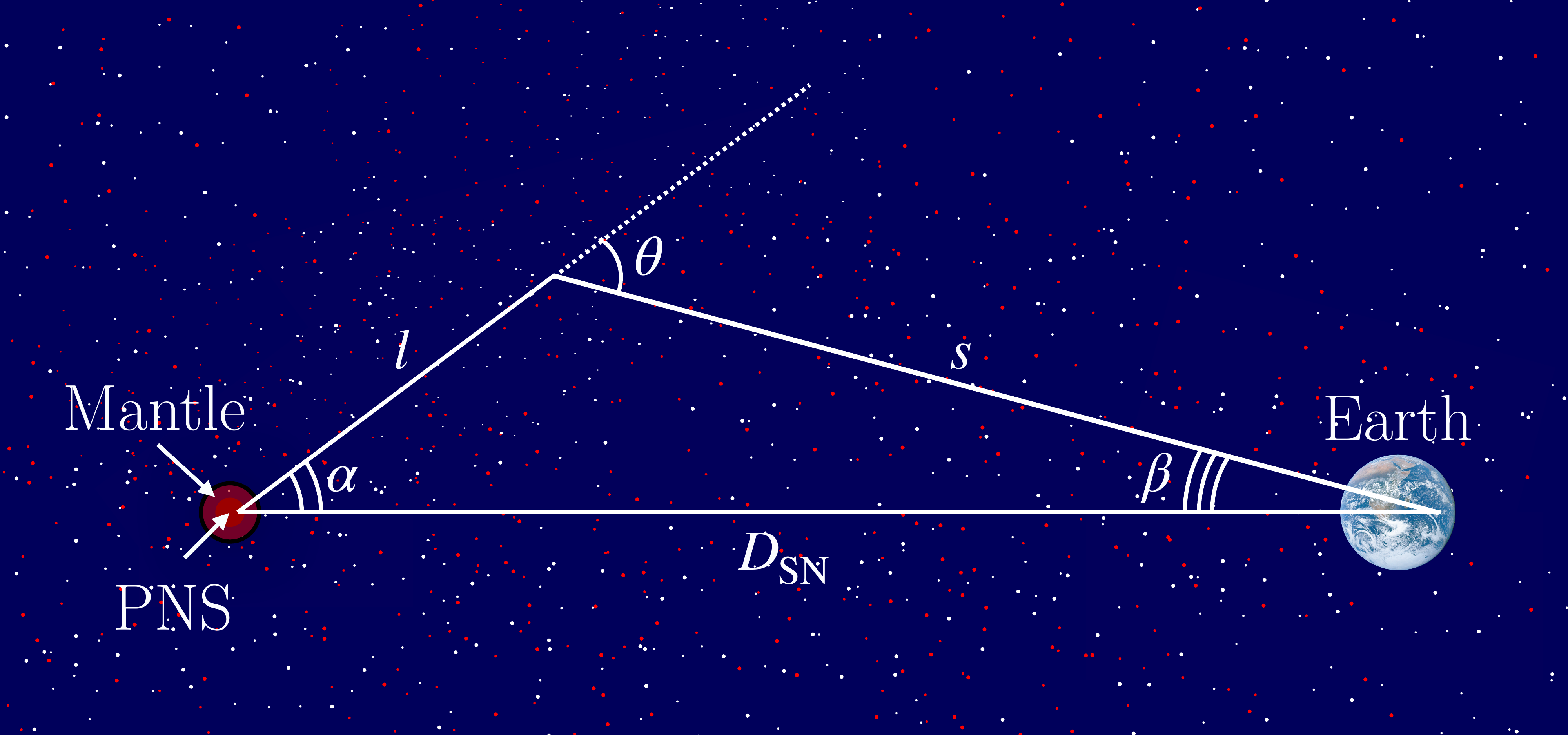}
    \caption{\textbf{\textit{Difference between the time of arrival of neutrinos sourced by the neutrinosphere and by Majoron decay.}}
    The length travelled by a neutrino emitted by the PNS surface is $D_{\rm SN}$, while a neutrino produced through Majoron decay accumulates a delay. The resulting time spread of the signal depends on the Majoron coupling to neutrinos, $g$, but not on the Majoron mass, $m_\phi$ (see Appendix~\ref{app:time_distribution}); figure is not to scale, since for our parameter space $l\ll D_{\rm SN}$. Combining timing and energy information one can obtain the mass and coupling of the Majoron to neutrinos.}
    \label{fig:drawing}
\end{figure}

As observed in \Refe~\cite{Akita:2022etk}, the next galactic core-collapse SN constitutes a preferred target for future searches of novel particles that couple to neutrinos. Upcoming neutrino detectors will be orders-of-magnitude larger than present ones, and the latest large dark-matter detectors will also be able to observe neutrinos from SNe. In this paper, we revisit the discovery reach of Majorons by combining new detectors and showing how the results depend on the assumed SN model. Crucially, we show that combining all the information on the time-dependent (rather than the time-integrated) neutrino flux at Earth, i.e., energy, flavor, and time of arrival of the neutrinos, we might be able to partially reconstruct the couplings and mass of $\phi$. First, the flavor information, which can be uncovered from the different event topologies in the detector, can be directly connected with the individual flavor couplings $g_\alpha$.
Second, although the time-integrated flux depends on the combination $g_{\alpha}m_\phi$, the high-energy neutrinos produced by Majoron decays would arrive spread in time with respect to the standard SN neutrinos (see Fig.~\ref{fig:drawing}), with the width of the spread depending on the couplings only, thus breaking the degeneracy between couplings and mass. While we obtain our results explicitly for the simple interaction Lagrangian of Eq.~\eqref{eq:Lagrangian}, mutatis mutandis one can generalize our findings to any BSM scenario featuring a particle light enough to be produced by SNe and that can decay to neutrinos.

\section{Neutrinos from Majoron decay}
\label{sec:neutrinos_from_Majorons}

In the SN core, neutrinos and anti-neutrinos have thermal potentials of equal size but opposite signs, dominated by neutrons, with absolute value $|V|=G_{\rm F} \rho/\sqrt{2}m_N\sim 10$~eV, where $m_N$ is the nucleon mass, for a typical nuclear density $\rho\sim 3\times 10^{14}$~g~cm$^{-3}$. For a typical neutrino energy $E_\nu\sim 100$~MeV, if the Majoron mass  $m_\phi \gtrsim \sqrt{|V| E_\nu}\sim 10$~keV, the dominant Majoron production channel is the coalescence of pairs of neutrinos, $\nu_\alpha + \nu_\alpha \to\phi$, and anti-neutrinos, $\bar{\nu}_\alpha + \bar{\nu}_\alpha \to\phi$.  Summing over neutrinos and anti-neutrinos of all flavors ($\alpha = e, \mu, \tau$), the resulting energy spectrum of emitted Majorons of energy $E_\phi$, per unit volume and time, is~\cite{Fiorillo:2022cdq}
\begin{equation}\label{eq:ndot}
    \frac{d\dot{n}_\phi}{dE_\phi}=\sum_{\sigma=\pm}\sum_{\alpha}\frac{g_{\alpha}^2 m_\phi^2}{64\pi^3}\int_{E_-}^{E_+}\!\!dE_\nu\, f^\sigma_\alpha (E_\nu) f^\sigma_\alpha (E_\phi-E_\nu) \;,
\end{equation}
where $f^\sigma_\alpha(E_\nu)=\bigl[e^{(E_\nu-\sigma\mu_\alpha)/T}+1\bigr]^{-1}$ is the Fermi-Dirac distribution of $\nu_\alpha$ (for $\sigma = +1$) or $\bar{\nu}_\alpha$ (for $\sigma = -1$), which are in local thermal equilibrium with temperature $T$ and have chemical potential $\sigma\mu_\alpha$, $E_{\pm} \equiv \frac{1}{2}(E_\phi\pm p_\phi)$, and $p_\phi=\sqrt{E_\phi^2-m_\phi^2}$ is the momentum of the Majoron.  Local thermal equilibrium holds throughout the proto-neutron star (PNS), where the bulk of the Majoron emission occurs. Because inside the PNS electron neutrinos are degenerate, with a chemical potential of $\mu_e \sim 100$~MeV, they dominate the Majoron production and thus the typical energy of the Majorons is comparable, i.e., $E_\phi \sim \mu_e$. 

The total number of Majorons produced throughout the SN evolution is
\begin{equation}\label{eq:number_emitted_Majorons}
    \frac{dN_\phi}{dE_\phi}=\int \frac{d\dot{n}_\phi}{dE_\phi} dV dt \; ,
\end{equation}
where the integral is over the spatial and temporal profiles of a supernova simulation. We use two one-dimensional supernova models, SFHo-18.8 and LS220-s20.0, that were evolved using the {\sc Prometheus Vertex} code with six-species neutrino transport~\cite{JankaWeb}.  They represent, respectively, the case of a very heavy (final neutron star baryonic mass of $1.926\;M_\odot$, where $M_\odot$ is one solar mass) and a very light PNS (final neutron star baryonic mass of $1.351\;M_\odot$). The profiles of temperature and chemical potential for both models are shown in the Supplemental Material of \Refe~\cite{Fiorillo:2022cdq}. Following \Refes~\cite{Bollig:2020xdr, Caputo:2021rux,Fiorillo:2022cdq,Fiorillo:2024upk}, we dub them the hot and cold models, respectively, as they present comparably high (up to $60$~MeV) and low temperatures (up to $40$~MeV). We extract the chemical potentials for $\nu_e$ and $\nu_\mu$ from the simulations. For $\nu_\tau$, for which the chemical potential is not available, we consider zero chemical potential. In reality, for the latter, a small chemical potential does build up, because $\overline{\nu}_\tau$ interact less with nucleons than $\nu_\tau$ due to weak-magnetism corrections to the cross sections~\cite{Horowitz:2001xf}, but remains significantly smaller than for the other neutrino species.\footnote{In Ref.~\cite{Fiorillo:2022cdq} the Majoron production from $\nu_\tau$ was assumed to be equal to the one from $\nu_\mu$ in the SN core. For the results of Ref.~\cite{Fiorillo:2022cdq}, based on flavor-universal coupling, this makes little to no difference, since the production from $\nu_\mu$ is anyway very subdominant with respect to $\nu_e$ in the cold model, and even in the hot model, which produced the aggressive bounds, it only changes the Majoron luminosity by some $30\%$, leading to a $10\%$ change in the bounds on the coupling. However, for flavor-dependent coupling as we consider here, the difference can be sizable and must be considered.
}

The Majoron energy emission is in general time-dependent.  Its time structure, which is shown in the Supplemental Material of Ref.~\cite{Fiorillo:2022cdq}, peaks immediately after the bounce, since it is correlated with the large chemical potential of neutrinos. However, for our analyses we always consider the time-integrated emission of Majorons, Eq.~\eqref{eq:number_emitted_Majorons}.

After Majorons are emitted, they decay via $\phi \to \nu_\alpha + \nu_\alpha$ at a rest-frame rate of
\begin{equation}
    \Gamma_{\phi\to\nu\nu}=\sum_\alpha\frac{g_{\alpha}^2 m_\phi}{32\pi} \;,
\end{equation}
and at an identical rate via $\phi \to \bar{\nu}_\alpha + \bar{\nu}_\alpha$; hence, their total decay rate is $2\Gamma_{\phi\to\nu\nu}$. In the laboratory frame, the Majoron lifetime is lengthened by a factor $E_\phi/m_\phi$ due to time dilation. Nevertheless, for the ranges of couplings and masses that we consider, the Majorons decay long before reaching Earth. Even for a coupling of $gm_\phi\sim 10^{-11}$~MeV, the decay length of a Majoron with energy $E_\phi\sim 100$~MeV is about $10^{-3}$~pc, tiny compared to the 10-kpc-scale distances to the SN that we consider. The daughter neutrinos have energies comparable to the parent Majoron, i.e., in the $100$-MeV range, significantly higher than standard SN neutrinos (Section~\ref{sec:standard_sn_neutrinos}), whose energies are in the 10-MeV range. Thus, the detection of high-energy neutrinos at Earth would provide an identifying signature of Majorons.

Majorons decay into $\nu_\alpha$ or $\bar{\nu}_\alpha$ with branching ratios $\mathrm{Br}(\phi\to\nu_\alpha+\nu_\alpha)=\mathrm{Br}(\phi\to\bar{\nu}_\alpha+\bar{\nu}_\alpha)=g_\alpha^2/(2\sum_\beta g_\beta^2)$; the factor of $2$ accounts for the decay in neutrinos and anti-neutrinos evenly. As neutrinos propagate, they mix; upon reaching Earth, the fraction of $\nu_\alpha$ in the total flux is $f_\alpha=\sum_{i=1,2,3} \sum_{\beta=e,\mu,\tau}|U_{\alpha i}|^2 |U_{\beta i}|^2 \mathrm{Br}(\phi\to\nu_\beta+\nu_\beta)$, where $U$ is the Pontecorvo-Maki-Nakagawa-Sakata (PMNS) matrix.
The energy spectrum of the daughter neutrinos from the decay of Majorons has a characteristic box shape flat in neutrino energy, so the energy fluence of $\nu_\alpha$ at Earth is
\begin{equation}\label{eq:n-nu}
    \frac{d\mathcal{F}_\alpha}{dE_\nu}=\frac{2f_\alpha}{4\pi D_\mathrm{SN}^2}
    \int_{E_{\rm min}}^{\infty}\frac{dE_\phi}{p_\phi}\frac{dN_\phi}{dE_\phi}\Big|_{E_\phi} \;,
\end{equation}
and identically for $\overline{\nu}_\alpha$, where $D_\mathrm{SN}$ is the SN distance from Earth; the factor $2$ accounts for the two daughter neutrinos produced in each decay.  We use $D_\mathrm{SN}=10$~kpc as a reference value, corresponding roughly to a SN occurring at the galactic center. The minimum Majoron energy needed to produce a neutrino of energy $E_\nu$ is $E_{\rm min}=E_\nu+m_\phi^2/(4E_\nu)$. Setting $f_\alpha=1/6$ in Eq.~\eqref{eq:n-nu} reproduces the analogous result of \Refe~\cite{Fiorillo:2022cdq}. The spectrum of the neutrinos emitted from the Majoron decay is shown in Fig.~2 of \Refe~\cite{Fiorillo:2022cdq}.

For sufficiently small couplings, $g_\alpha$, and large masses, $m_\phi$, the Majorons would be significantly slowed down compared to the standard neutrinos. The typical time scale after which a Majoron decays is roughly $\tau_\phi\sim E_\phi/(\Gamma_{\phi\to\nu\nu}m_\phi)\sim E_\phi/\sum_\alpha g_\alpha^2 m_\phi^2$. Within this time scale, the Majoron accumulates a delay due to its velocity, $v_\phi=\sqrt{1-m_\phi^2/E_\phi^2}$, of about $\tau_\phi (1-v_\phi)\sim 1/\sum_\alpha g_\alpha^2 E_\phi$. The daughter neutrinos are produced at a typical angle $\theta\sim m_\phi^2/E_\phi^2$ with respect to the direction of the Majoron, so this angular deflection produces a comparable delay.  The net result is that the neutrino emission from Majoron decay is not delayed relative to the emission of standard neutrinos, since some of the Majorons decay early, but rather stretched over a time scale of about $\tau_\phi (1-v_\phi)$, which is independent of the Majoron mass,  as it would be for any BSM model with an operator of dimension~4.
We account for the above effects by computing the time-dependent neutrino energy spectrum at Earth as
\begin{equation}\label{eq:ndot-nu}
    \frac{d\Phi_\alpha(t)}{dE_\nu}=\frac{d\mathcal{F}_\alpha}{dE_\nu} \exp\left[-\frac{2\Gamma_{\phi}E_\nu t}{m_\phi}\right]\frac{2\Gamma_\phi E_\nu}{m_\phi} \;.
\end{equation}
In Eq.~\eqref{eq:ndot-nu}, while the spectrum of daughter neutrinos, $d\mathcal{F}_\alpha/dE_\nu$, depends on the Majoron couplings and mass---roughly $\propto g_\alpha m_\phi$, for low values of $m_\phi$---the terms introduced by the stretching of their emission time depend only on the coupling.  In our time-dependent analysis (Section~\ref{sec:time_dependent_flavor_insensitive}) this will break the degeneracy between coupling and mass.
We review the derivation of Eq.~\eqref{eq:ndot-nu} in Appendix~\ref{app:time_distribution}; see also, e.g., \Refe~\cite{Oberauer:1993yr}.

\section{Standard supernova neutrinos}\label{sec:standard_sn_neutrinos}

The vast majority of the neutrinos from the next galactic core-collapse SN will be made in SM processes. 
For the purpose of our work, they act as background to our search for the neutrinos made in the decay of Majorons.  
Here we review the main properties of the standard SN neutrino flux and discuss how we model it. For a more detailed discussion of SN neutrino emission, we refer to dedicated reviews, e.g., \Refe~\cite{Janka:2012wk}.

The SN explosion is triggered by the collapse of the core of a very massive star. When the core reaches nuclear densities, the collapse is suddenly halted and a shock wave is launched towards the outer stellar layers. Neutrinos in the inner regions are kept in thermal and chemical equilibrium by weak interactions with the surrounding nuclear matter, with $\nu_e$ being highly degenerate due to the relatively large lepton number trapped in the SN core. The region where neutrinos remain in equilibrium with matter is approximately bounded by a neutrinosphere, beyond which the neutrino interaction rate with matter drops sufficiently for neutrinos to escape; this happens typically at a distance of about $20$~km from the center of the core. In reality, the concept of neutrinosphere is a fuzzy one, and different species decouple at different distances, but it provides a qualitative grasp on the distances involved.

At the earliest stage of the SN, neutrino emission is due to the shock breakout from the neutrinosphere, leading to a burst---the neutronization burst---of $\nu_e$ produced in the neutronization of free protons $p + e^- \to n + \nu_e$ in the core. This burst has a high instantaneous luminosity, of the order of $10^{53}$~erg~s$^{-1}$, but it is short, lasting only about 50~ms. In our analysis, we do not consider neutrinos from the neutronization burst. Rather, our focus is on the longer-duration neutrino emission of later SN stages, and on the high-energy neutrinos created in the decays of Majorons, altogether spanning tens of seconds.

After the shock breakout, the matter behind the shock starts to accrete onto the surface of the PNS, powering neutrino emission.  During this accretion phase, lasting 0.5--1~s, the shock wave remains essentially stalled. The PNS undergoes sustained neutrino emission with typical luminosities of $10^{52}$~erg~s$^{-1}$. In the neutrino delayed shock mechanism, a fraction of this energy is deposited below the shock wave, reviving it and leading to the SN explosion. Afterwards, the PNS undergoes a cooling phase, where neutrinos are emitted with a quasi-thermal spectrum, for a typical duration of $5$--$6\,\rm s$.  At later times, additional emission might come from fallback accretion of material, an intrinsically three-dimensional process which is known with less certainty than the previous phases.

The total energy of the emitted neutrinos, $E_{\rm tot}$, is roughly $10^{53}$~erg, and has a marked dependence on the mass of the PNS, with heavier PNSs yielding higher neutrino luminosities; the equation of state can also impact the neutrino emission.
The average energy of the emitted neutrinos, $E_0$, is typically 12--14~MeV. Following \Refes~\cite{Keil:2002in, Tamborra:2012ac, Vitagliano:2019yzm}, we model the time-integrated (excluding the neutronization burst), all-species energy spectrum of the emitted neutrinos as
\begin{equation}\label{eq:nufromSN}
    \frac{dN_{\nu}}{dE_\nu}=\frac{E_{\mathrm{tot}}}{E_0^2} \frac{(1+\alpha)^{1+\alpha}}{\Gamma(1+\alpha)} \left(\frac{E_\nu}{E_0}\right)^\alpha e^{-(1+\alpha)E_\nu/E_0} \;,
\end{equation}
where $\Gamma$ is the Gamma function and $\alpha$ is the pinch parameter; $\alpha=2$ corresponds to a Maxwell-Boltzmann spectrum. 
The emitted spectra are different for the different species, with $\nu_e$ having lower energies than the other ones due to their higher interaction rate, which keeps them in equilibrium with nuclear matter at larger radii with lower temperatures. However, the differences are generally small, with the average energy of $\nu_e$ typically around $9$~MeV and of the other species typically around $12$~MeV. The total emitted energy for different species is quite similar, with up to $\sim 5\%$ differences (see, e.g., Table~VII of \Refe~\cite{Fiorillo:2023frv}).  
We ignore these small differences and assume a common shape, Eq.~\eqref{eq:nufromSN}, for the spectrum of all neutrino species. For the two SN models that we consider, we use the values of the parameters in Eq.~\eqref{eq:nufromSN} from \Refe~\cite{Fiorillo:2022cdq}: in the cold model, $E_{\rm tot} = 1.98\times 10^{53}$~erg, $E_0 = 12.7$~MeV, and $\alpha = 2.39$; in the hot model, $E_{\rm tot} = 3.93\times 10^{53}$~erg, $E_0 = 14.3$~MeV, and $\alpha = 2.07$. 

The flavor composition of the neutrino flux detected at Earth, i.e., the proportion of $\nu_e$, $\nu_\mu$, and $\nu_\tau$ in it, is affected by neutrino mixing and its interplay with weak interactions. 
In the regions close to the neutrinosphere, weak interactions can induce fast flavor conversions~\cite{Sawyer:2004ai, Sawyer:2005jk, Sawyer:2015dsa, Chakraborty:2016lct, Izaguirre:2016gsx, Tamborra:2020cul, Volpe:2023met,Fiorillo:2024bzm} where neutrinos traveling along different directions change flavor via the refractive forward scattering $\bar{\nu}_e + \nu_e\to \bar{\nu}_\mu + \nu_\mu$.
Although the kinetic equations describing fast flavor conversions are well-known~\cite{Dolgov:1980cq, Rudsky, Sigl:1993ctk, Sirera:1998ia, Yamada:2000za, Vlasenko:2013fja, Volpe:2013uxl, Serreau:2014cfa, Kartavtsev:2015eva,Fiorillo:2024wej,Fiorillo:2024fnl}, their outcome and impact on supernova evolution largely is not; see, e.g., \Refes~\cite{Tamborra:2020cul,Capozzi:2022slf} and references therein.
The typical length scale over which flavor would change is of the order of cm, much shorter compared to the hydrodynamical scale, making a direct numerical implementation of both scales a formidable problem beyond current computational capabilities. When the dynamics is entirely driven by the refractive exchange, numerical~\cite{Nagakura:2022kic} and theoretical~\cite{Fiorillo:2024qbl} arguments suggest that flavor equipartition is reached along certain directions, but the interplay between the refractive exchange and neutrino collisions with matter remains an open question~\cite{Shalgar:2020wcx, Martin:2021xyl, Sigl:2021tmj, Sasaki:2021zld, Johns:2021qby, Padilla-Gay:2022wck, Hansen:2022xza, Johns:2022yqy, Lin:2022dek, Xiong:2022zqz, Fiorillo:2023ajs}. In any case, the feedback of the altered neutrino flavor composition on the hydrodynamical simulations has not been self-consistently studied, although it appears that if large changes occur there could be considerable impact on the supernova evolution~\cite{Ehring:2023lcd, Ehring:2023abs}.

In the outer regions of the SN, where the neutrino density is lower, the interplay between neutrino mixing in vacuum and refractive flavor exchange could still induce slow flavor conversions, which could give rise to energy-dependent flavor swaps~\cite{Duan:2006an,Raffelt:2007xt,Raffelt:2007cb}. In addition, standard Mikheyev–Smirnov–Wolfenstein (MSW) resonances~\cite{Wolfenstein:1977ue, Mikheyev:1985zog} in neutrino-matter interactions would affect neutrino mixing, but their precise effect depends on the unknown degree of adiabaticity of the matter density profile and the ordering of the neutrino mass eigenstates, and thus remains unclear; for details, see \Refes~\cite{Duan:2010bg,Capozzi:2022slf} and references therein.  In any case, predictions of these behaviors in the outer regions are unavoidably conditioned by the lack of understanding of fast conversions.

In view of the above uncertainties, we restrict ourselves to the simplest assumption of species equipartition in the standard SN neutrino flux at Earth, \ie, that the flux is evenly divided among $\nu_e$, $\bar{\nu}_e$, $\nu_\mu$, $\bar{\nu}_\mu$, $\nu_\tau$, and $\bar{\nu}_\tau$ but point out that this should be revisited later, once flavor conversions inside SNe are better understood.  These uncertainties do not affect Majoron emission, which happens in the core, where neutrinos are in chemical equilibrium with the surrounding matter and neutrino mixing is inhibited.
 
As pointed out above, the standard neutrino flux acts as a background to our search for the high-energy neutrinos from Majoron decays. Since the abundance of the latter peaks in the $100$-MeV range, there is a clear separation between the two neutrino signals, and our sensitivity to neutrinos from Majoron decays comes mainly from these high energies.  This fact renders the impact of how we model the standard SN signal on our results largely unimportant. 
In the energy window between 30--70~MeV, the overlap between the standard and non-standard SN components does require that the two are disentangled. Hence, we model the time-integrated SN neutrino fluence with the spectrum from Eq.~\eqref{eq:nufromSN}, assuming, as mentioned above, equipartition among different species. Relieving these analysis choices would enlarge the model parameter space of our analysis without adding any new qualitative feature to it, nor affecting our results sizably, since they rest rather on the starkly different energy scales of the standard and non-standard neutrino fluxes. 

We only model the time-integrated neutrino fluence. In principle, for a time-dependent analysis like the one we perform later (Section~\ref{sec:analysis}), it would be helpful to also disentangle the standard neutrino flux from the flux of neutrinos from Majoron decays by using some parametrization of the time-dependent standard flux. However, no such parametrization presently exists. A first attempt in this direction has only recently been performed in \Refe~\cite{Lucente:2024ngp}; however, it is only for the time dependence of the energy-integrated neutrino luminosity, not for the energy spectrum. Thus, when we do turn to a time-dependent analysis, we choose conservatively to simply remove the low-energy events in order to eliminate the SN neutrino flux rather than model it, as we discuss in detail in Section~\ref{sec:analysis}.

\section{Neutrino detection from the next galactic supernova}\label{sec:neutrino_detection}

The next galactic supernova will lead to the detection of a very large signal at a host of different observatories. Here, we are mainly interested in the possibility that a small fraction of high-energy events might be observed, which would be a smoking-gun signature of novel bosons produced in the SN core such as the Majoron we study here. If these events were to be observed, we aim to understand to what degree they could be used to reconstruct the properties of the novel bosons.  Below, we review the main detection channels of SN neutrinos and estimate their expected detection rate.

\subsection{Neutrino detection processes}
\label{sec:neutrino_detection-processes}

All of the neutrino detection processes entail the scattering of a neutrino or an anti-neutrino off a nucleus, a nucleon, or an electron.  We list below the main processes.

\subsubsection{Quasi-elastic (QE) electron neutrino scattering}

In water Cherenkov detectors, $\overline{\nu}_e$ can be detected via inverse beta decay (IBD) on free protons,~$\bar{\nu}_e + p \to n + e^+$, and both $\nu_e$ and $\overline{\nu}_e$ through their charged-current scattering on oxygen nuclei, \mbox{$\nu_e + \ce{^16O} \to e^- + X$} and $\bar{\nu}_e + \ce{^16O} \to e^+ + Y$, where $X$ and $Y$ denote final excited nuclear states. For $\bar{\nu}_e$, the latter reactions dominate over IBD at neutrino energies larger than about 70 MeV. Above the kinematic threshold, the cross section for QE neutrino--nucleus ($\nu A$) scattering grows as $\sigma_{\nu_e A}\propto E_\nu^2$. Therefore, a high-energy neutrino from Majoron decay is more likely to be detected than a lower-energy standard SN neutrino. We neglect the recoil energy of the nucleus, so that the energy of the final-state electron is simply $E_{e}=E_{\nu}-Q_{\nu_e A}$, and $E_{e}=E_{\nu}-Q_{\bar{\nu}_e A}$ for the positron, where the transferred momentum, $Q_{\nu_e A}$ or $Q_{\bar{\nu}_e A}$, depends on the type of nucleus; specifically, $Q_{\bar{\nu}_e p}=1.29$~MeV, $Q_{\bar{\nu}_e O}=11.4$~MeV, and $Q_{\nu_e O}=15.4$~MeV. Therefore, the differential number of electrons produced by $\nu_e A$ interactions is
\begin{equation}
   \label{eq:det_rate_qel_e}
   \frac{dN_{e^-}}{dE_e dt}
   =
   \sum_{A=p,O} 
   N_{A,\mathrm{tg}} \sigma_{\nu_e A}
   (E_e+Q_{\nu_e A}) 
   \left. \frac{d\Phi_\alpha}{dE_\nu} \right\vert_{E_\nu=E_e+Q_{\nu_e A}} \;,
\end{equation}
where $N_{A,\mathrm{tg}}$ is the number of target nuclei $A$.  The expression for the number of positrons produced by $\bar{\nu}_e A$ interactions, $dN_{e^+}/dE_e dt$, is analogous, with $\sigma_{\nu_e A} \to \sigma_{\bar{\nu}_e A}$ and $Q_{\nu_e A} \to Q_{\bar{\nu}_e A}$.
We use the total QE neutrino--nucleus cross sections, $\sigma_{\nu_e A}$ and $\sigma_{\bar{\nu}_e A}$, from 
\Refes~\cite{Marteau:1999zp,Kolbe:2002gk,Strumia:2003zx,Formaggio:2012cpf}, collected in Fig.~S1 of \Refe~\cite{Fiorillo:2022cdq}.

\subsubsection{Quasi-elastic (QE) muon neutrino scattering}

Muon neutrinos and anti-neutrinos are detected via reactions analogous to the ones above, but with a different kinematic threshold. For $\bar{\nu}_\mu$ with energies larger than $E_{\nu, \textrm{thres}} \equiv m_\mu + m_n - m_p\simeq 107$~MeV, where $m_\mu$, $m_n$, and $m_p$ are, respectively, the masses of the muon, neutron, and proton, the reaction $\bar{\nu}_\mu + p \to \mu^+ + n$ is kinematically allowed~\cite{Formaggio:2012cpf}. While this reaction is practically irrelevant for standard SN neutrinos because of the threshold, it is important for neutrinos from Majoron decays. The charged-current reactions  $\nu_\mu + \ce{^16O} \to \mu^- + X$ and $\bar{\nu}_\mu + \ce{^16O} \to \mu^+ + Y$ are likewise relevant above a similar energy threshold. Like before, the  cross section scales as $\sigma_{\nu_\mu A} \propto E_\nu^2$. 
Again, we neglect the nucleus recoil and write the energy of the final-state muon as $E_{\mu}=E_{\nu}-Q_{\nu_\mu A}$, and similarly for the final-state anti-muon.  Because the transferred momenta depend only on the nucleus properties, they are the same for $\nu_\mu$ and $\bar{\nu}_\mu$ as for $\nu_e$ and $\bar{\nu}_e$, i.e., $Q_{\nu_\mu A} = Q_{\nu_e A}$ and $Q_{\bar{\nu}_\mu A} = Q_{\bar{\nu}_e A}$. In analogy to Eq.~\eqref{eq:det_rate_qel_e}, the differential number of muons produced by $\nu_\mu A$ interactions is
\begin{equation}
   \label{eq:det_rate_qel_mu}
   \frac{dN_{\mu^-}}{dE_\mu dt}
   =
   \sum_A N_{A,\mathrm{tg}} 
   \sigma_{\nu_\mu A}(E_\mu+Q_{\nu_\mu A}) 
   \left. \frac{d\Phi_\alpha}{dE_\nu} \right\vert_{E_\nu=E_\mu+Q_{\nu_\mu A}} \;,
\end{equation}
and analogously for the number of anti-muons from $\bar{\nu}_\mu A$ interactions, $dN_{\mu^+}/dE_\mu dt$. We use the total cross sections, $\sigma_{\nu_\mu A}$ and $\sigma_{\bar{\nu}_\mu A}$, from Refs.~\cite{Formaggio:2012cpf, Marteau:1999zp}, collected in Fig.~S1 of \Refe~\cite{Fiorillo:2022cdq}.  

Final-state muons with energies above $E_{\mu,\mathrm{Cher}}=160 \,\rm MeV$ produce Cherenkov radiation in water. These ``visible muons'' can be identified in water Cherenkov detectors. Their detection rate, based off of Eq.~\eqref{eq:det_rate_qel_mu}, is
\begin{equation}
    \label{eq:det_rate_visible_muons}
    \frac{dN^\mathrm{vis}_{\mu^\pm}}{dE_\mu dt}
    =
    \frac{dN_{\mu^\pm}}{dE_\mu dt}
    \Theta(E_\mu-E_{\mu,\mathrm{Cher}}) \;,
\end{equation}
where $\Theta$ is the Heaviside theta function.
    
In contrast, lower-energy muons do not emit Cherenkov radiation. Instead, they stop quickly due to energy losses by ionization.   
These ``invisible muons'' decay, at rest, into detectable electrons and positrons whose energy spectrum is the Michel spectrum,
\begin{equation}
    \frac{dN_{\mu\to e}}{dE_e}=\frac{4}{m_\mu}\left(\frac{2E_e}{m_\mu}\right)^2\left(3-\frac{4E_e}{m_\mu}\right) \;,
\end{equation}
normalized here to yield one electron or positron from the decay of one muon or anti-muon. The spectrum has a sharp cut-off at  $E_e = m_\mu/2$. Thus, the differential number of electrons and positrons produced from invisible muons is
\begin{equation}
    \label{eq:det_rate_invisible_muons}
    \frac{dN_{e^\pm}^\mathrm{inv}}{dE_e dt}=\frac{dN_{\mu\to e}}{dE_e}\int^{E_{\mu,\mathrm{Cher}}}_{m_\mu}dE_\mu\frac{dN_{\mu^\pm}}{dE_\mu dt} \;.
\end{equation}
The Michel electrons and positrons carry no memory of the original $\nu_\mu$ and $\bar{\nu}_\mu$ spectrum. Invisible muons cannot be separated on an event-by-event basis from other types of events that are also detected via final-state electrons and positrons, but their Michel spectrum, with its characteristic cut-off, can be identified via the statistical analysis we perform below, providing some information on the flavor composition of high-energy neutrinos.

\subsubsection{Neutrino-electron elastic scattering (ES)}

In addition to the above QE reactions, neutrinos and anti-neutrinos of all flavors can undergo ES on electrons, $\nu_\alpha + e^- \to \nu_\alpha + e^-$. The cross section for this channel, $\sigma_{\nu_\alpha e}$, differs for neutrinos of different flavors---it is largest for $\nu_e$---but is generally much smaller than the QE neutrino-nucleus cross sections, and grows instead as $\sigma_{\nu_\alpha e} \propto m_e E_\nu$. This is insufficient to compensate the rapidly falling neutrino flux at high energies (Fig.~\ref{fig:flavor_coupling_reconstruction_benchmark}). Nevertheless, we do include it, as it provides some sensitivity to all the different flavors, differently from the QE neutrino-nucleus scatterings and especially from IBD. 

The final-state electrons in ES are emitted in the forward direction, since the scattered electron is much lighter than the average neutrino energies. In contrast, electrons from QE neutrino-nucleus scatterings are emitted isotropically due to the large nucleus mass. Photons from the de-excitation of the final-state nuclei in QE scattering, as well as the time delay to neutron capture in IBD interactions, improve the tagging of these types of events~\cite{Scholberg:2012id}. Thus, we consider ES events as a separate topology, distinguishable from the others by selecting events that point back to the direction of the SN from whence they come.

The rate of $\nu_\alpha + e \to \nu_\alpha + e$ interactions in the detector is
\begin{equation}
   \label{eq:det_rate_elastic}
   \frac{dN_e}{dE_e dt}
   =
   N_{e,\mathrm{tg}} 
   \int dE_\nu 
   \frac{d\Phi_{\nu_\alpha}}{dE_\nu} 
   \frac{d\sigma_{\nu_\alpha e}}{dE_e} \;,
\end{equation}
where $N_{e,\mathrm{tg}}$ is the number of target electrons.
The expression for the rate of $\bar{\nu}_\alpha + e \to \bar{\nu}_\alpha + e$ interactions is analogous, with $d\sigma_{\nu_\alpha e}/dE_e \to d\sigma_{\bar{\nu}_\alpha e}/dE_e$.  We adopt the cross sections, $d\sigma_{\nu_\alpha e}/dE_e$ and $d\sigma_{\bar{\nu}_\alpha e}/dE_e$ valid for $E_\nu > 5$~MeV, from Eqs.~(10.17)--(10.18) of \Refe~\cite{Raffelt:1996wa}.  In principle, for the high-energy neutrinos that we target, the reaction $\nu_\mu + e^- \to \nu_e + \mu^-$ might also be accessible if $E_\nu > m_\mu \approx 105$~MeV. However, in analogy to $\nu_e e$ ES, the $\nu_\mu e$ ES  rate above 100~MeV is negligible, so we do not consider it.
    
\subsubsection{Coherent elastic neutrino-nucleus scattering}

Detectors sensitive to low-energy nuclear recoils can measure coherent elastic neutrino-nucleus scattering (CE$\nu$NS), $\nu + A \to \nu +  A$~\cite{Freedman:1973yd, COHERENT:2017ipa, CONUS:2020skt, Abdullah:2022zue}. The cross section for this interaction, $\sigma_{\textrm{CE}\nu\textrm{NS}}$, is flavor-blind at tree level and $\sigma_{\textrm{CE}\nu\textrm{NS}}\propto N_n^2$, where $N_n$ is the number of neutrons in the nucleus $A$. 
Thus, one could wonder whether the flavor-blind contribution from CE$\nu$NS can complement the information from the other topologies, sensitive only to the electron and muon flavors, to pinpoint the flavor composition of the BSM neutrinos. We find that the rate of CE$\nu$NS is very small in the detector that we adopt (see, e.g., \figu{flavor_coupling_reconstruction_benchmark}) and that it does not substantially affect our results, but we include it in our calculations nevertheless.

The differential rate of recoil nuclei from CE$\nu$NS is
\begin{align}
   \label{eq:det_rate_coherent}
   \frac{dN_A}{dE_R dt} = N_{A, \mathrm{tg}} \int dE_\nu \frac{d\Phi_\nu}{dE_\nu}\frac{d\sigma_{\textrm{CE}\nu\textrm{NS}}}{dE_R}\ \;,
\end{align}
where $N_{A, \mathrm{tg}}$ is the number of target nuclei, and $E_R$ is the observed recoil energy. Here, $d\Phi_\nu/dE_\nu$ is the all-species flux.
We consider the cross section as in \Refe~\cite{Lang:2016zhv}. From the kinematics of the reaction, the maximum nuclear recoil energy for a neutrino of energy $E_\nu$ is
\begin{align}
   E^\textrm{max}_R = \frac{2E^2_\nu}{2E_\nu + m_A}\simeq\frac{2E^2_\nu}{m_A}\, ,
\end{align}
where $m_A$ is the mass of the target nucleus; for xenon, this is of the order of $100$~GeV. Hence, for neutrinos with energies of tens of MeV, the recoil energies are typically in the keV range.  Only scatterings with momentum transfer lower than about 50 MeV are coherent on nuclei~\cite{Scholberg:2020pjn}, as enforced by the nuclear form factors embedded in the CE$\nu$NS cross section. While there are significant uncertainties in the form factors for large momentum transfer~\cite{Papoulias:2019lfi}, we do not worry about them, given the small impact of CE$\nu$NS on our results.

\medskip

In conclusion, for our analysis, we consider four distinct detection channels. QE electron neutrino scattering combined with the Michel spectrum (dubbed ``QE $e^\pm +$ Inv. $\mu^\pm$'') primarily provide information about the $\nu_e$ and $\nu_\mu$ content of the Majoron neutrino fluxes. Visible $\mu^\pm$ from QE muon neutrino scattering inform our analysis about the $\nu_\mu$ content alone. ES and CE$\nu$NS are used as independent probes, providing information about all flavors.

\subsection{Experimental facilities}

A collection of next-generation neutrino observatories are currently under construction or being planned. These are generally multi-purpose detectors with a broad physics program. Here, we comment on the neutrino detectors considered in our analysis---Hyper-Kamiokande and DARWIN---and the potential contributions of other experiments.

\subsubsection{Hyper-Kamiokande}

Hyper-Kamiokande~\cite{Hyper-Kamiokande:2018ofw} is the planned successor of Kamiokande and Super-Kamiokande. As a water Cherenkov detector, it is suitable for the study of neutrinos with energies from few MeV to hundreds of GeV. In this work, we consider an experimental configuration of two tanks with a fiducial volume of 187~kton each.\footnote{This is the fiducial volume with a background-reducing veto. For a SN, one could in principle use the full volume of 220~kton per tank, although the results would only marginally be affected.} This corresponds to a total number of {$2.5\times 10^{34}$} proton targets, {$1.25\times 10^{34}$} $\ce{^16O}$ targets, and {$1.25\times 10^{35}$} electrons. In our analysis, we include events from the QE scattering of electron and muon neutrinos and anti-neutrinos, and from ES on electrons separately.

\subsubsection{DARWIN}

DARWIN~\cite{Aalbers:2022dzr,DARWIN:2016hyl} will be a dark matter and neutrino detector consisting of 30 tonnes of xenon, corresponding to {$1.8\times 10^{29}$} target nuclei, for which we assume an isotopic composition following that of its natural abundance. It will be sensitive to {CE$\mathbf{\nu}$NS}, although, as pointed out above, the detection rate for the BSM signal would be too low to help. DARWIN will also be sensitive to elastic neutrino-electron scattering, but we do not consider this detection channel in it, since the interaction rates at Hyper-Kamiokande will dominate.

\subsubsection{Other experiments}
\label{sec:neutrino_detection-experiments_other}

Our analysis is focused on Hyper-Kamiokande and DARWIN.  Although other neutrino detectors might be in operation by the time the next galactic SN explodes, these two detectors, together, encompass all the detection channels listed above.  Below, we list a few other detectors that could observe the SN, though we do not consider them in our work.

The European Spallation Source Neutrino Beam (ESS$\nu$B)~\cite{ESSnuSB:2023ogw} is a planned Cherenkov detector consisting of two enormous water tanks that act as far detectors. They are comparable in size to Hyper-Kamiokande and so could increase significantly the number of detected events via IBD, QE neutrino-nucleus scattering, and neutrino-electron scattering. If built, ESS$\nu$B would increase the effective volume for Hyper-Kamiokande and consequently improve our results; however, we do not include it in our current analysis.

Liquid-scintillator detectors like the Jiangmen Underground Neutrino Observatory (JUNO)~\cite{JUNO:2015zny, JUNO:2021vlw} are sensitive to $\bar{\nu}_e$ via IBD, to neutrino ES on electrons, and to QE scattering on carbon, in analogy to water Cherenkov detectors. However, because JUNO is roughly twenty times smaller than Hyper-Kamiokande, its contribution to the detection rate would be unimportant in comparison to it.

Although primarily conceived as a long-baseline neutrino oscillation experiment, the planned Deep Underground Neutrino Experiment (DUNE) can also detect neutrinos with energies above 1~MeV~\cite{DUNE:2020zfm, DUNE:2020ypp}. DUNE will detect neutrinos and anti-neutrinos via their charged-current and neutral-current interactions on liquid argon, and will observe the electron neutrino component of the flux via $\nu_e + \ce{^40Ar} \to e^- + \ce{^40K}^*$. However, despite the relatively large cross section of this process, the limited fiducial volume of DUNE would yield a number of events much lower than the number of IBD events detected by Hyper-Kamiokande.

Besides DARWIN, other CE$\nu$NS detectors, like LZ~\cite{Khaitan:2018wnf}, PandaX~\cite{Pang:2024bmg}, PICO~\cite{Kozynets:2018dfo}, and RES-NOVA~\cite{RES-NOVA:2021gqp} might be able to observe a few more neutrinos from the decay of Majorons. Nonetheless, the comparatively small target mass of these detectors and the uncertainties in the nuclear form factors for large momentum transfer severely limit their sensitivity. 

The IceCube neutrino telescope, located at the South Pole, is a km-scale in-ice Cherenkov detector optimized for the study of neutrinos with energies above 100~GeV.  However, it is expected to be sensitive also to tens-of-MeV neutrinos from a galactic SN~\cite{Köpke_2011, IceCube:2011cwc}. For SN neutrinos, the large size of IceCube could, in principle, yield a large detection rate. However, IceCube cannot resolve individual neutrino events from a SN because their energies are too low. Instead, its conventional search strategy for SN neutrinos is to look for a detector-wide increase in the activity of the detector modules.

Looking for coincidences between modules~\cite{Klein:2013nbr,Fritz:2021btf} would allow one to reconstruct at least the average energy of the SN neutrino spectrum, but the latter would be largely dominated by the standard SN neutrinos. Thus, any information on the possible presence of high-energy BSM neutrinos would be lost. For a search like ours, that seeks to disentangle the high-energy, non-standard component from the low-energy, standard component, this is a serious limitation. Thus, we do not include IceCube among the experiments that we use in our work, since by itself it cannot provide a unique signature of the neutrinos from Majoron decay. In principle, should the data from other experiments---Hyper-Kamiokande in particular---reveal high-energy neutrinos, the IceCube observations could be combined with these data to enhance statistics. Such an analysis would not rely on any \textit{qualitatively} new observable, but it would require carefully combining the high-statistics, energy-blind information from IceCube with the low-statistics, energy-dependent information from Hyper-Kamiokande. Hence, we deem it more instructive to perform it with real data, should the next galactic SN reveal hints of new physics, and do not include it in our projections.

Recently, \Refe~\cite{Lazar:2024ovc} proposed to use the time structure of the high-statistics signal at IceCube to probe either \textit{i)} light Majorons ($m_\phi \lesssim 10$~MeV) whose decay produces neutrinos peaking at bounce time, rather than at $0.1\,\rm s$ as the standard signal does, or \textit{ii)} heavy Majorons ($m_\phi \approx 200$~MeV) producing neutrinos with a substantial stretch in time ($\sim 10\,\rm s$). We believe that in both cases the use of IceCube alone cannot really probe these scenarios, since it cannot separate the high-energy non-standard neutrinos from the lower-energy standard neutrinos. 

For case \textit{i)}, the signature proposed in \Refe~\cite{Lazar:2024ovc} is an excess within the first $0.1$~s. However, for the size of the Majoron couplings that are considered, $g_\phi \sim 10^{-10}$, the daughter neutrinos from their decay would induce an excess of only about $500$~counts within the first $0.1$~s, whereas the standard flux would produce about $10^4$~counts in the same time window (numbers extracted from Fig.~3 of \Refe~\cite{Lazar:2024ovc}). This excess is only about $5\%$ of the overall signal. To unambiguously claim that such an excess is non-standard \textit{without} knowing the energy of the individual events requires a comparable systematic uncertainty on the theoretical SN models of 1--10\%. The uncertainty due to the unknown equation of state and PNS mass, as well as the impact of many other microphysical aspects (\eg,  convection, the presence of muons, the modeling of neutrino--nucleon interaction rates) is much larger (see, \eg, the simulations in \Refe~\cite{Fiorillo:2023frv} and related fits in \Refe~\cite{Lucente:2024ngp}). Therefore, it seems unlikely that such an excess would truly allow us to identify new physics.  Observations made by other experiments will likely provide some information to reconstruct the standard SN signal, but this means that meaningful conclusions regarding neutrinos from Majoron decays involving IceCube could be drawn only from a combined analysis of different detectors, accounting for both energy and timing information. We do not expect such analysis to improve on the results of Hyper-Kamiokande alone, since an $\mathcal{O}(1)$ excess of events at Hyper-Kamiokande corresponds approximately to a fraction $10^{-4}$ of the number of standard SN events, much smaller than the 1--10\% fraction that can be constrained by IceCube.

For case \textit{ii)}, the time window that seems to dominate the projected IceCube bounds, identified in Fig.~3 of Ref.~\cite{Lazar:2024ovc}, is between $0.05$~s and $11$~s. The ratio between the signal and the background is still of the order of 1-10\%, so our comments for case $\textit{i})$ apply. In addition, this case also depends on the standard signal at late times, a phase which is quite uncertain. While the cooling phase lasts up to $7$~s, it is entirely possible that standard neutrinos are emitted over many tens of seconds, driven by the late-time accretion of stellar material onto the PNS; see, \eg, \Refes~\cite{1989ApJ...346..847C, Janka1996, Zhang:2007nw, Sukhbold:2015wba, Ott:2017kxl, Vartanyan:2019ssu, Burrows:2019zce, Li:2020ujl}. In fact, a recent reanalysis of the SN~1987A observations~\cite{Fiorillo:2023frv} shows that fallback accretion after 9--10~s is a likely explanation for the late-time events observed at Kamiokande and BUST, which would otherwise be challenging to explain within a pure PNS cooling picture.

Aside from the above issues, the projected bounds on the Majoron coupling from Hyper-Kamiokande  in \Refe~\cite{Lazar:2024ovc} are a factor-of-4 weaker than the bounds we obtain ourselves in Section~\ref{sec:projected_bounds} (which in turn agree with those in \Refe~\cite{Akita:2022etk} to within about 40\%, as well as with our order-of-magnitude estimates).  This discrepancy corresponds to a factor-of-16 difference in event rates. It originates from \Refe~\cite{Lazar:2024ovc} assuming an atmospheric background over a full day~\cite{LazarPrivateComm}. In reality, the SN neutrino burst lasts only 10~seconds, and even for relatively heavy Majorons with masses $m_\phi\sim 100$~MeV, for the smallest couplings that can be constrained at Hyper-Kamiokande (see Fig.~\ref{fig:projected_bounds}), the spread in the signal reaches at most $10^3$~s. Considering a full day of exposure ($86400$~s) for the atmospheric flux overestimates the background by 2--4 orders of magnitude. Using a background rate for a single tank of Hyper-Kamiokande of $55$~events per day (re-scaled from the $2$~events per day expected at IMB to the $187$~kton volume of Hyper-Kamiokande) we find that even in $10^3$~s, less than one background event is expected. For light Majorons, the relevant duration is closer to $1$--$10$~s, with an even smaller background. So, the search should be considered as background-free, justifying our neglect of atmospheric neutrinos in this work, akin to what was done in \Refe~\cite{Akita:2023iwq}. The Hyper-Kamiokande reach obtained in this work is stronger than the reach of IceCube and IceCube-Gen2, which was, in any case, obtained assuming perfect knowledge of the SN signal.

\subsection{Expected event rates}\label{sec:expected_event_rates}

To compute the expected number of detected events, we use the predicted non-standard (Section~\ref{sec:neutrinos_from_Majorons}) and standard (Section~\ref{sec:standard_sn_neutrinos})  SN neutrino flux, and the interaction rates of the different detection channels we consider (Section~\ref{sec:neutrino_detection}).
   We account for the uncertainty in the energy of detected events by using a Gaussian energy resolution function centered on the true value  $E_{\rm tr}$ of the final-state electron, positron, muon, or anti-muon,
   \begin{equation}\label{eq:gaussian_resolution}
    \mathcal{R}(E_{\rm tr}, E_{\rm rec}) = 
    \frac{1}{\sqrt{2\pi} \delta(E_{\rm tr})} \mathrm{exp}\left(-\frac{(E_{\rm tr}- E_{\rm rec})^2}{2\delta(E_{\rm tr})^2}\right) \;,
   \end{equation}
   where $\delta = c_1 E_{\rm tr} + c_2 \sqrt{E_{\rm tr}} +c_3$ is the width of the resolution and $E_{\rm rec}$ is the reconstructed energy.  Table \ref{tab:resolution_constants} contains the values of constants used for each detection channel. Thus, the event rate as a function of the reconstructed energy is
   \begin{equation}
    \label{eq:det_rate_rec}
    \frac{dN}{dE_\mathrm{rec}dt}=\int dE_\mathrm{tr} \frac{dN}{dE_\mathrm{tr}dt}\mathcal{R}(E_\mathrm{tr},E_\mathrm{rec}) \;,
   \end{equation}
   where $dN/dE_{\rm tr}dt$ is either Eq.~\eqref{eq:det_rate_qel_e}, \eqref{eq:det_rate_visible_muons}, \eqref{eq:det_rate_invisible_muons}, \eqref{eq:det_rate_elastic}, or \eqref{eq:det_rate_coherent}. 
   We assume perfect detection efficiency, given the lack of a precise estimate of the attainable efficiency. In reality, a lower efficiency could be compensated by a larger detector volume.
   In any case, it is not expected to significantly affect the region of parameter space probed, given that the event rate is proportional to the square of the coupling, which is therefore weakly sensitive to order-unity changes in the rate. 

\begin{table}[t!]
    \caption{\textbf{\textit{Parameters of the energy resolution function in neutrino detection.}} The function, $\delta$, is defined immediately below Eq.~\eqref{eq:gaussian_resolution}.}
    \renewcommand{\arraystretch}{1.4}
    \centering
    \begin{tabular}{cc|ccc|c}
        \hline
        Channel & Detector & $c_1$  & $c_2$ [MeV$^{1/2}$] & $c_3$ [MeV] & Ref.\\\hline 
        QE $e^\pm$ + Inv. $\mu^\pm$ & Kamiokande & 0 & 0.866 & 0 & \cite{Jegerlehner:1996kx}\\
        QE $e^\pm$ + Inv. $\mu^\pm$ & IMB & 0 & 1.16 & 0 & \cite{Jegerlehner:1996kx}\\\hline
        QE $e^\pm$ + Inv. $\mu^\pm$ & Hyper-Kamiokande & 0 & 0.1 & 0 & \cite{Martinez-Mirave:2024hfd}\\
        Visible $\mu^\pm$ & Hyper-Kamiokande & 0 & 0.1 & 0& \cite{Martinez-Mirave:2024hfd}\\
        Electron scattering & Hyper-Kamiokande & 0.0397 & 0.349 & -0.0839 & \cite{Super-Kamiokande:2016yck}\\
        CE$\nu$NS & DARWIN & 7.7$\times 10 ^{-2}$ & 7.3$\times 10^{-3}$ & 6.9$\times 10 ^{-5}$ & \cite{Schumann:2015cpa}\\ \hline
    \end{tabular}
    \label{tab:resolution_constants}
\end{table}

\section{Analysis of the neutrino signal from the next galactic supernova}\label{sec:analysis}

Neutrino detection has come a long way since the observation of SN~1987A. The current and imminent neutrino and dark matter detectors will collect enough data from the next galactic SN to either place much more powerful bounds on BSM particles coupling to neutrinos, or potentially to discover an additional flux component on top of the standard SN neutrinos. The number of detected events could be large enough to probe the flavor-dependent couplings of a Majorons to neutrinos (Section~\ref{sec:introduction}). We explore both possibilities by identifying the regions of the mass and coupling parameter space that will become accessible when the next galactic supernova is detected. For our projections, unless otherwise specified, we stick to the cold SN model---but marginalize over the cold and hot model in our statistical treatment---meaning that if a hotter SN were to explode, the results could be much better due to the larger number of detected neutrinos.

\begin{figure}
    \centering
    \includegraphics[width=0.95\textwidth]{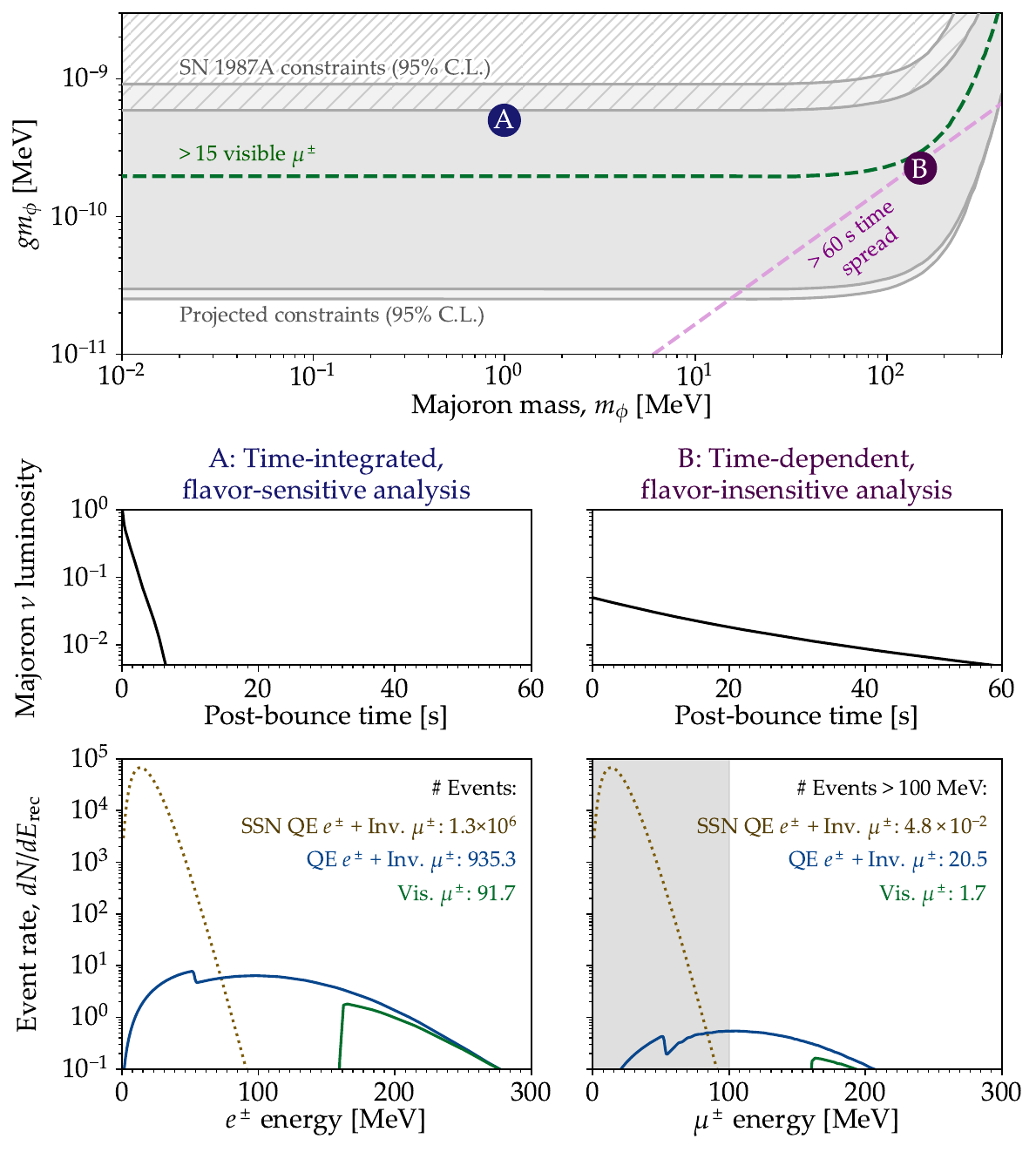}
    \caption{\textit{\textbf{A schematic of our analysis in the Majoron--neutrino coupling parameter space.}} \textit{Top panel:} The plane $g m_\phi$--$m_\phi$ of flavor-universal coupling, $g$, and Majoron mass, $m_\phi$, showing current SN~1987A and projected flavor-universal constraints. Flavor-dependent coupling analysis is possible with a high-enough Visible $\mu^\pm$ event rate, illustrated with the contour showing $> 15$ visible muons. A mass-reconstruction analysis using timing information is possible with a long-enough time spread, illustrated with the contour showing a $> 60$~s time spread. The lower panels show the timing (\textit{middle}) and energy (\textit{bottom}) distributions of neutrinos predicted for the time-independent, flavor-sensitive benchmark (left), and the time-dependent, flavor-insensitive benchmark (right). For the energy spectrum, we also show results for the Standard SN (SSN) neutrinos, exclusively for the QE $e^\pm$ + Inv. $\mu^\pm$ signal.}
    \label{fig:overview}
\end{figure}

Figure \ref{fig:overview} sketches our analysis, which is two-fold, depending on the size of the couplings:

\begin{description}
 \item[Time-integrated, flavor-sensitive analysis] 
  For large enough couplings, large samples of detected events will be available that could allow us not only to detect the high-energy, non-standard signal from Majoron decays, but also to reconstruct, at least in part, the flavor structure of the coupling, i.e., to measure the values of the couplings of the different neutrino flavors, $g_e$, $g_\mu$, and $g_\tau$, individually. Because, for such large couplings, the typical time spread of the neutrinos from Majoron decay is short compared to the duration of the standard SN neutrino signal, we perform a time-integrated analysis. Our sensitivity is driven by the interplay of the QE $e^\pm$ + Inv.~$\mu^\pm$ and visible $\mu^\pm$ channels, which are sensitive to different flavors (Section~\ref{sec:neutrino_detection}).
  In \figu{overview}, we identify coarsely the region in the parameter space of Majoron mass and coupling (flavor-universal coupling, for illustration) where such a time-integrated, flavor-sensitive analysis is viable due to there being more than $15$~high-energy detected visible muons from Majoron decay. 
 \item[Time-dependent, flavor-insensitive analysis]
  For a sufficiently small value of the coupling $g$, the Majoron time spread estimated in Section~\ref{sec:neutrinos_from_Majorons} can become larger than the standard duration of the burst. While the number of detected events is lower in this region of the parameter space, the time structure of the detected neutrino signal would provide a direct probe of the size of the coupling, whereas the number of detected events would depend primarily on the combination $g m_\phi$ of coupling and mass. We will show that this allows for a clear disentangling of the two parameters and, therefore, an explicit reconstruction of the Majoron mass. In Fig.~\ref{fig:overview}, this region of the parameter space is coarsely identified by the requirement that the typical time spread, $m_\phi/2\Gamma_\phi E_\nu$, is larger than $60$~s for a 100-MeV neutrino.
\end{description}

When real data become available, a complete time-dependent analysis with flavor sensitivity might provide more detailed information. Today, however, this poses serious technical challenges, since there is no flexible parametrization for the time- and energy-dependent signal from a SN. Nevertheless, our approach needs not face these challenges, instead using the conservative strategy based on the separation between the energies of the standard and non-standard neutrinos.

\subsection{Time-integrated, flavor-sensitive analysis}

\subsubsection{Bounds from the non-observation of high-energy neutrinos}
\label{sec:projected_bounds}

Before asking the question of how well one could reconstruct the Majoron properties in case a high-energy SN neutrino signal were observed, we deal with the more pessimistic scenario in which no such signal is observed. In this case, the next galactic SN will still provide novel bounds on the Majoron coupling more powerful than the current ones.  

Previously, \Refes~\cite{Akita:2023iwq, Lazar:2024ovc} reported projected bounds from a galactic SN. Here, we independently compute projected bounds. We make projections assuming both the cold and hot models, while \Refe~\cite{Akita:2023iwq} used the cold model only and \Refe~\cite{Lazar:2024ovc} used a model of electron-capture SN with a very low progenitor mass~\cite{Hudepohl:2009tyy}, using an equation of state nowadays not strongly favored~\cite{JankaPrivateComm}; apart from this, such a model should give results quite close to the cold model given the similar PNS mass. Further, \Refes~\cite{Akita:2023iwq, Lazar:2024ovc} considered only the QE $e^\pm$ + Inv. $\mu^\pm$ topology, although this should not affect too much the bounds which are sensitive to the total number of non-standard events, largely dominated by this class of events.

\bigskip
\noindent
\textit{Bounds from SN~1987A}
\medskip

First, we obtain present-day bounds on the Majoron mass and coupling using the SN~1987A observations. These bounds were obtained in \Refe~\cite{Fiorillo:2022cdq} for the case of $g_e=g_\mu=g_\tau$. We follow closely the same procedure, but extend it to the case where $g_e$, $g_\mu$, and $g_\tau$ can be different from one another. For the bounds based on the legacy data from SN~1987A we use only the $e^\pm$ signal from QE scattering detected by Kamiokande and IMB, and neglect the visible muons due to the uncertainties in how the detectors would have seen such a signal. 

We briefly summarize the statistical procedure here, emphasizing the few differences with \Refe~\cite{Fiorillo:2022cdq}.  We use the unbinned likelihood function
\begin{align}    
 \label{eq:likelihood_bounds_sn1987a}
 \mathcal{L}\left(g_\alpha,m_\phi;\boldsymbol{\theta}_\mathrm{SN},\boldsymbol{\theta}_\mathrm{mix}\right)
 = \;  
 & \mathrm{exp}\left(
 -\int_{E_{\rm min}}^{E_{\rm max}} dE_{\rm rec}
 \frac{d N_{\rm QE}}{d E_{\rm rec}}
 \right)
 \prod_k \frac{d N_{\rm QE}}{d E_{\rm rec}} (E_k)
 \nonumber\\
 & \times \pi_{12}(\theta_{12})\pi_{13}(\theta_{13})\pi_{23,{\rm CP}}(\theta_{23},\delta_{\rm CP}) \;,
\end{align}
where $E_{\rm rec}$ is the reconstructed energy of the event, ${d N_{\rm QE}}/{d E_{\rm rec}}$ is the event rate in the QE $e^\pm$ + Inv. $\mu^\pm$ detection channel, Eqs.~\eqref{eq:det_rate_qel_e}, \eqref{eq:det_rate_invisible_muons}, and \eqref{eq:det_rate_rec}, and
$E_k$ is the observed energy of each event at Kamiokande and IMB. The minimum energy $E_{\rm min}$ is chosen for each experiment following the prescription in \Refe~\cite{Fiorillo:2022cdq}, chosen to minimize the contamination from background. The maximum energy $E_{\rm max}=600$~MeV, but the results are independent of this value since the signal peaks at approximately $150$~MeV. In the likelihood, we make explicit the dependence on the Majoron couplings and mass, $g_\alpha$ and $m_\phi$, the parameters of the standard SN flux [Eq.~\eqref{eq:nufromSN}], $\boldsymbol{\theta}_\mathrm{SN}=\left\{E_\mathrm{tot},E_0,\alpha\right\}$, and the neutrino mixing parameters that parametrize the PMNS matrix~\cite{ParticleDataGroup:2022pth}, $\boldsymbol{\theta}_\mathrm{mix}=\left\{\theta_{12},\theta_{13},\theta_{23},\delta_{\rm CP}\right\}$. Differently from \Refe~\cite{Fiorillo:2022cdq}, we also introduce pull factors,
$\pi$, that evaluate the probability distribution function (PDF) of the mixing parameters.  For these, we adopt the $\chi^2$ distributions from the NuFit~5.2 global fit to oscillation data, assuming normal neutrino mass ordering~\cite{Esteban:2020cvm, NuFit5.2}. For the mixing parameters $\theta_{12}$ and $\theta_{13}$, the PDFs are one-dimensional; for $\theta_{23}$ and $\delta_{\rm CP}$, the joint PDF is two-dimensional, since they are highly correlated. 
In the case of flavor-universal couplings studied in \Refe~\cite{Fiorillo:2022cdq}, the neutrino flux at Earth is independent of the mixing matrix, so the pull factors were inessential and not used. In our analysis, this is no longer true.

To obtain the bounds, for each trial value of $g_\alpha$ and $m_\phi$ we compute the profiled likelihood $\overline{\mathcal{L}}(g_\alpha,m_\phi) = \mathrm{max}_{\theta_\mathrm{SN},\theta_{\mathrm{mix}}} \mathcal{L} $. The one exception is that we do not profile over the value of the pinch parameter, $\alpha$, since the sparse data of SN~1987A are essentially insensitive to it~\cite{Mirizzi:2005tg,Fiorillo:2023frv}. Instead, we use the fixed value of $\alpha$ for the hot and cold SN models reported in Section~\ref{sec:standard_sn_neutrinos}. 
For each value of $m_\phi$, the likelihood is maximum for vanishing values of $g_\alpha$, confirming that the SN~1987A detection did not provide any hint of Majorons. Therefore, we define the test statistic (TS), $\mathrm{TS}(g_\alpha,m_\phi)=-2\,\left(\log\left[\overline{\mathcal{L}}(g_\alpha,m_\phi)\right]-\log\left[\mathrm{max}_{g_\alpha}(\overline{\mathcal{L}}(g_{\alpha},m_{\phi}))\right]\right)$. Under the null hypothesis that a Majoron with parameters $g_\alpha$ and $m_\phi$ exists, for each value of $m_\phi$ the value of TS should be distributed as a half-$\chi^2$ distribution (i.e., $\chi^2/2$) with one degree of freedom~\cite{Cowan:2010js}.\footnote{More specifically, the TS should be distributed as a half-$\chi^2$ distribution only if its definition is slightly modified, such that if $g_\alpha<\hat{g}_\alpha$, where $\hat{g}_\alpha$ is the best-fit value of the coupling, it is defined equal to 0. However, since for our data $\hat{g}_\alpha=0$, this modified definition has no practical effect.} To obtain the bounds, we set the $95\%$ confidence level (C.L.) threshold value, $\mathrm{TS}(g_\alpha,m_\phi)=2.7$. 

\begin{figure}
    \centering
    \includegraphics[width=0.95\textwidth]{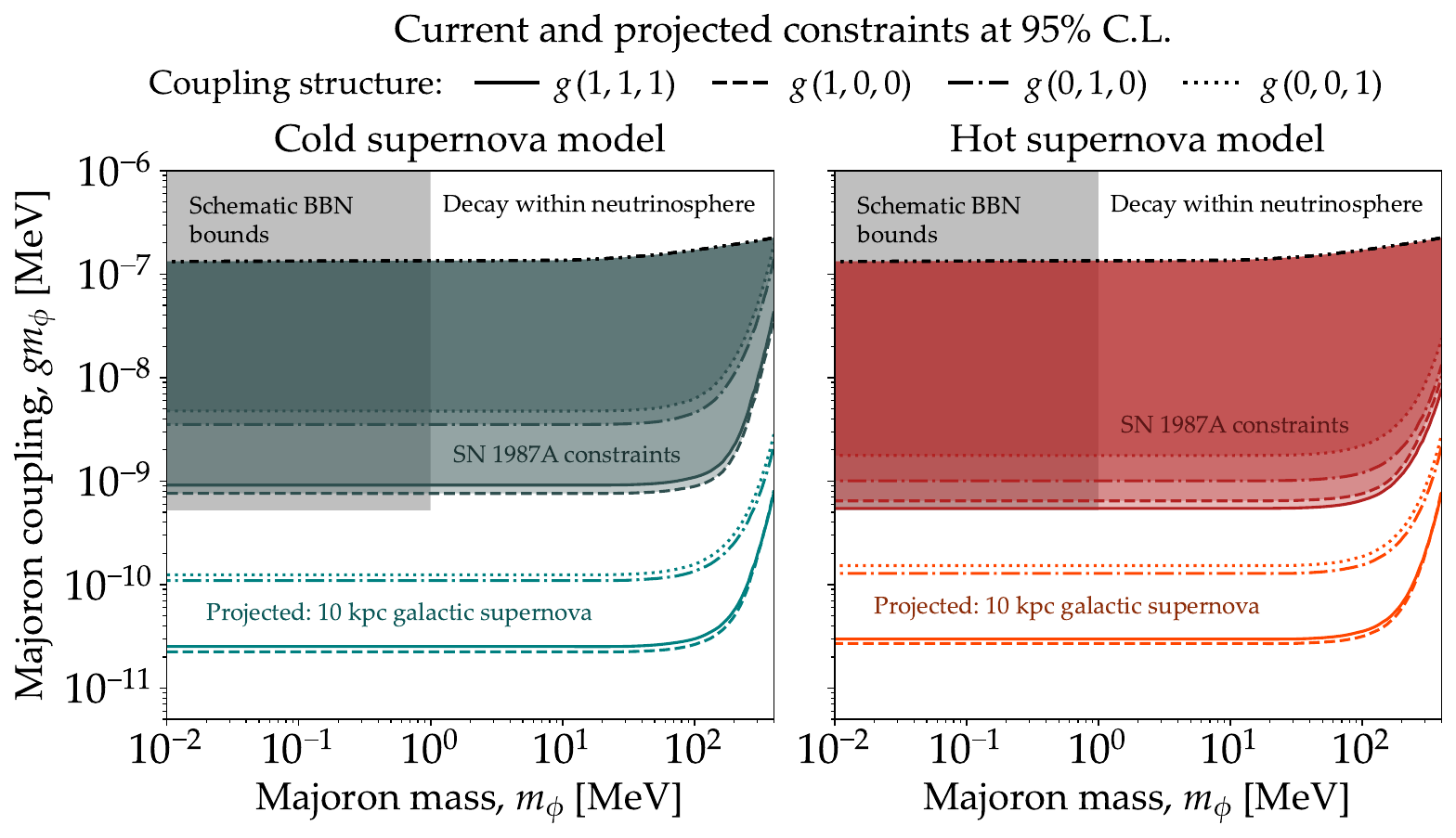}
    \caption{\textit{\textbf{Majoron mass and coupling constraints from SN~1987A data and from a projected next galactic SN at 10~kpc.}} The schematic BBN bounds are taken from \cite{Fiorillo:2022cdq}. Above $g m_\phi\sim 10^{-7}$~MeV, our bounds do not apply due to trapping since Majorons decay within the neutrinosphere. The SN~1987A bounds are obtained \textit{assuming} that the SN was cold (\textit{left}) or hot (\textit{right}); hence, they are conservative and aggressive. The projected bounds are obtained assuming that the next galactic SN will be cold (\textit{left}) or hot (\textit{right}), but in both cases we do not assume knowledge of the model in the analysis; hence, the bounds are pessimistic and optimistic but in both cases conservative.
    }
    \label{fig:projected_bounds}
\end{figure}

Figure~\ref{fig:projected_bounds} shows the bounds we obtain for four benchmark cases of the flavor texture of the couplings: universal flavor couplings ($g_e = g_\mu = g_\tau$, as in \Refe~\cite{Fiorillo:2022cdq}), $\nu_e$-only coupling ($g_e \neq 0$, $g_\mu = g_\tau = 0$), $\nu_\mu$-only coupling ($g_\mu \neq 0$, $g_e = g_\tau = 0$), and $\nu_\tau$-only coupling ($g_\tau \neq 0$, $g_e = g_\mu = 0$). We show bounds separately under the assumptions that the neutrinos from Majoron decay are produced in the cold and hot SN models. The bounds made in the cold and hot model, respectively, should then be interpreted as being conservative and aggressive. 

For the flavor-universal case, we reproduce the results of \Refe~\cite{Fiorillo:2022cdq}, as expected, since we follow the same procedure, up to the anticipated $10\%$ differences induced by the different treatment of Majoron production from $\nu_\tau$. For the cold model, the Majorons produced from $\nu_e\nu_e$ coalescence are by far the dominant ones. Hence, the case $g_e\neq 0$ alone leads to bounds that are even more restrictive than the flavor-universal case, because the rate of Majoron production is mostly unchanged but the flavor composition in the decay contains more $\overline{\nu}_e$ at Earth, which contribute to the largest detection channel, IBD. For the same reason, the cases $g_\mu \neq 0$ and $g_\tau \neq 0$ lead to significantly weaker bounds, since they are worsened both by the much lower Majoron production and fraction of $\overline{\nu}_e$ at Earth. For the hot model, the Majoron production from $\nu_\mu \nu_\mu$ coalescence is nearly comparable with the production from $\nu_e \nu_e$ coalescence, due to the high temperatures in the SN core that lead to the build-up of a degenerate $\overline{\nu}_\mu$ population. Therefore, the bounds for $g_\mu \neq 0$ and $g_\tau \neq 0$ are quite closer to the bounds for $g_e\neq 0$. 

Reference~\cite{Akita:2023iwq} obtained bounds on the Majoron couplings akin to ours, for the cold model only, and with some differences. The bounds are comparable (see Fig.~5 in \Refe~\cite{Akita:2023iwq}), but ours are weaker by about $40\%$, implying a difference of nearly a factor of $2$ in the event rate. Since for SN~1987A we use the same strategy as \Refe~\cite{Fiorillo:2022cdq}, which \Refe~\cite{Akita:2023iwq} also follows, this difference is probably not due to the analysis procedure. When computing  Majoron production, \Refe~\cite{Akita:2023iwq} used the neutrino chemical potentials from a digitized version of the contour plots in \Refe~\cite{Fiorillo:2022cdq} which likely introduced imprecision in their values that led to the discrepancy in the bounds~\cite{AkitaPrivateComm}.
\newpage

\bigskip
\noindent
\textit{Projected bounds}
\medskip

Having obtained the present-day bounds, we can now obtain projected ones. We consider all the event topologies introduced earlier (Section~\ref{sec:neutrino_detection-processes}) to compute the event rates following the procedure in Section~\ref{sec:expected_event_rates}. We build a mock event sample by simulating the response of the detector to a hypothetical next galactic SN located at a distance $d=10$~kpc. We assume the next galactic SN to match the cold or the hot model in turn. In \textit{simulating} the signal we are not being conservative or aggressive, since this choice does not determine the strategy of analysis; in the analysis, we are always conservative and assume no knowledge of the simulated SN model. Our choice only determines whether the next galactic SN will be a cold or a hot one, bracketing the range of possibilities.

For each of the simulated SN models, we consider only the Asimov data sample, namely, that the experiment will detect a signal exactly equal to the expected one~\cite{Cowan:2010js}. In principle, the real signal might fluctuate around its mean, but the projected bounds are dominated by the high-energy region, above $100$~MeV, where the standard flux produces essentially no expected events and therefore fluctuations are irrelevant.
Even for our discovery prospects (Sections~\ref{sec:measuring_flavor_couplings} and \ref{sec:time_dependent_flavor_insensitive}), we neglect fluctuations away from the Asimov data sample, with the understanding that they may shift the best-fit values of the Majoron mass and coupling.

The expected value of the log-likelihood function, which is the quantity used to set the bounds (see, \eg, the discussion in Section~IV.B of \Refe~\cite{Fiorillo:2023clw}), is
\begin{equation}\label{eq:sim_likelihood}
\begin{split}
    \chi^2(g_\alpha,m_\phi)
    =
    & -2
    \, \mathrm{max}_{\boldsymbol{\theta}_\mathrm{SN}}\langle \log\mathcal{L}\left(g_\alpha,m_\phi; \boldsymbol{\theta}_{\rm SN}\right) \rangle
    \\ 
    =   
    & -2 \sum_{\rm t}
    \left[
    \int_0^{E_{\rm max}} dE_{\rm rec}\, \frac{dN_{\rm t,true}}{dE_{\rm rec}}\, \text{ln}\left(\frac{dN_{\rm t,test}(g_\alpha,m_\phi;\boldsymbol{\theta}_\mathrm{SN})}{dE_{\rm rec}}\right) 
    \right. 
    \\
    & \left. - \int_0^{E_{\rm max}} \, dE_{\rm rec}\, \frac{dN_{\rm t, test}(g_\alpha,m_\phi;\boldsymbol{\theta}_\mathrm{SN})}{dE_{\rm rec}}
    \right] \;,
\end{split}
\end{equation}
where $\langle \log \mathcal{L} \rangle$ is the likelihood function averaged over all possible realizations of the true distribution of the events---which corresponds to using the Asimov data sample---and $dN_{\rm t,test}/dE_{\rm rec}$ is the rate of events of topology t, evaluated at test values $g_\alpha$ and $m_\phi$ and test values of the SN parameters $\boldsymbol\theta_{\rm SN}$. The latter includes, as before, the parameters of the standard SN flux, $E_{\rm tot}$, $E_0$, and $\alpha$ (though $\alpha$ remains fixed, as for the SN 1987A bounds), but now also includes the choice between the hot and cold SN model, which we profile over when computing bounds. Similarly, $dN_{\rm t,true}/dE_{\rm rec}$ is the event rate evaluated at the assumed ``true'' values of the parameters.  In Eq.~\eqref{eq:sim_likelihood}, we set the maximum energy of integration to $E_{\rm max} = 600$~MeV for the QE $e^\pm$ + Inv.~$\mu^\pm$, visible $\mu^\pm$, and electron scattering channels, and to 10~eV for CE$\nu$NS; these choices cover the full width of the event energy distributions.  
Like  before, we define a test statistic,
\begin{equation}\label{eq:definition_ts}
    \mathrm{TS}(g_\alpha,m_\phi)=\chi^2(g_\alpha,m_\phi)-\chi^2(\hat{g}_\alpha,m_\phi) \;,
\end{equation}
where $\hat{g}_\alpha$ is the best-fit value of the coupling.  We compute constraints for varying values of $m_\phi$. Because we use Asimov data samples, the best-fit value of the coupling always coincides with the assumed true value.  We fix the mixing parameters to their present-day best-fit values from NuFit~5.2~\cite{Esteban:2020cvm, NuFit5.2}, assuming normal mass ordering, but ignore the future uncertainty on them, since we expect that it will be sizably reduced by upcoming oscillation experiments~\cite{Song:2020nfh}. Like for the SN~1987A bounds, we set the $95\%$~C.L.~threshold at ${\rm TS}(g_\alpha,m_\phi) = 2.7$.

To compute the projected bounds, we assume as true values $g_\alpha=0$. The set of simulated SN parameters, $\boldsymbol{\theta}_{\rm SN}$, corresponds to the choice between the cold or hot SN model. However, regardless of the choice of the true SN model, we always profile the likelihood over both the hot and cold models. In fact, profiling over the SN model in our analysis is perhaps an overly conservative step, since in reality the large standard SN signal would presumably allow us to infer quite precisely the mass of the PNS and to understand whether the SN was truly closer to a cold or a hot model. Thus, our projected bounds are conservative, since they do not assume knowledge of the SN model.

Figure~\ref{fig:projected_bounds} shows that the resulting projected bounds are a substantial improvement over the SN 1987A bounds. The improvement can be understood from simple order-of-magnitude estimates. Since the volume we use for Hyper-Kamiokande is $55$~times larger than the $6.8$~kton volume of IMB, we expect an improvement in the bounds of a factor of $\sqrt{55} \approx 7$. Our simulated SN is placed at a distance of $10$~kpc, shorter by a factor $5$ than SN~1987A, leading to an additional improvement of a factor of $5$. This totals an improvement of a factor $35$ in the bounds, in agreement with the results of our full numerical analysis in \figu{projected_bounds}. Our bounds differ from those of \Refe~\cite{Lazar:2024ovc}, as pointed out in Section~\ref{sec:neutrino_detection-experiments_other}.

\subsubsection{Measuring couplings to different neutrino flavors}
\label{sec:measuring_flavor_couplings}

If a sufficiently large high-energy neutrino flux is detected, the different detection channels allow for a partial reconstruction of the flavor structure of the Majoron couplings. To quantify this, we simulate the neutrino signal detected from the decay of a Majoron with $m_\phi=1$~MeV and $g m_\phi = 5 \times 10^{-10}$~MeV, which is currently not excluded, but lies near the edge of the most aggressive present-day bounds (\figu{projected_bounds}). For Majoron masses $m_\phi<10$~MeV, the event rate is essentially independent of the mass, so the results would be identical for any other value of the Majoron mass in this range. We adopt only the cold SN model as true, though we still profile the likelihood over the hot and cold models. This makes our projections pessimistic, since if a hotter SN exploded instead, we would expect higher event rates and more precise measurements.

In reconstructing the flavor structure of the couplings, we are faced with a Majoron parameter space including three diagonal couplings, $g_e$, $g_\mu$, and $g_\tau$, and the Majoron mass, $m_\phi$. In principle, there could be additional off-diagonal couplings of the form $g_{\alpha\beta}$, with $\alpha \neq \beta$, that violate lepton-flavor conservation; however, we choose to fix them to zero in our work [Eq.~\eqref{eq:Lagrangian}].  A comprehensive analysis of the entire parameter space would presumably be instructive to perform with real data. When making projections, such an analysis would only widen the parameter space and make the results less intuitive. Hence, we focus on a corner of the parameter space where $g_e$, $g_\mu$, or both are the only nonzero couplings, and we treat them as free parameters whose values we will determine. 

For the light Majorons that we consider for this analysis, with $m_\phi < 10$~MeV, the energy spectrum and time distribution of the neutrinos from their decay are essentially independent of the value of $m_\phi$ as discussed in Ref.~\cite{Fiorillo:2022cdq}, so we do not expect to have a strong sensitivity to its value.  In addition, discriminating between $g_\mu$ and $g_\tau$ is challenging,
since, because the mixing between them is near-maximal due to $\theta_{23} \approx 45^\circ$, they lead to similar flavor composition at Earth and, for the cold model, even similar Majoron production rates, as discussed in Section~\ref{sec:neutrinos_from_Majorons}. 
We simulate three benchmark cases, corresponding to the coupling structures $(g_e, g_\mu, g_\tau)= g\,(1,1,0)$, $g\,(1,0,0)$, and $g\,(0,1,0)$, where $g$ is the parameter whose value we float.

\begin{figure*}
    \centering
    \includegraphics[width=0.8\textwidth]{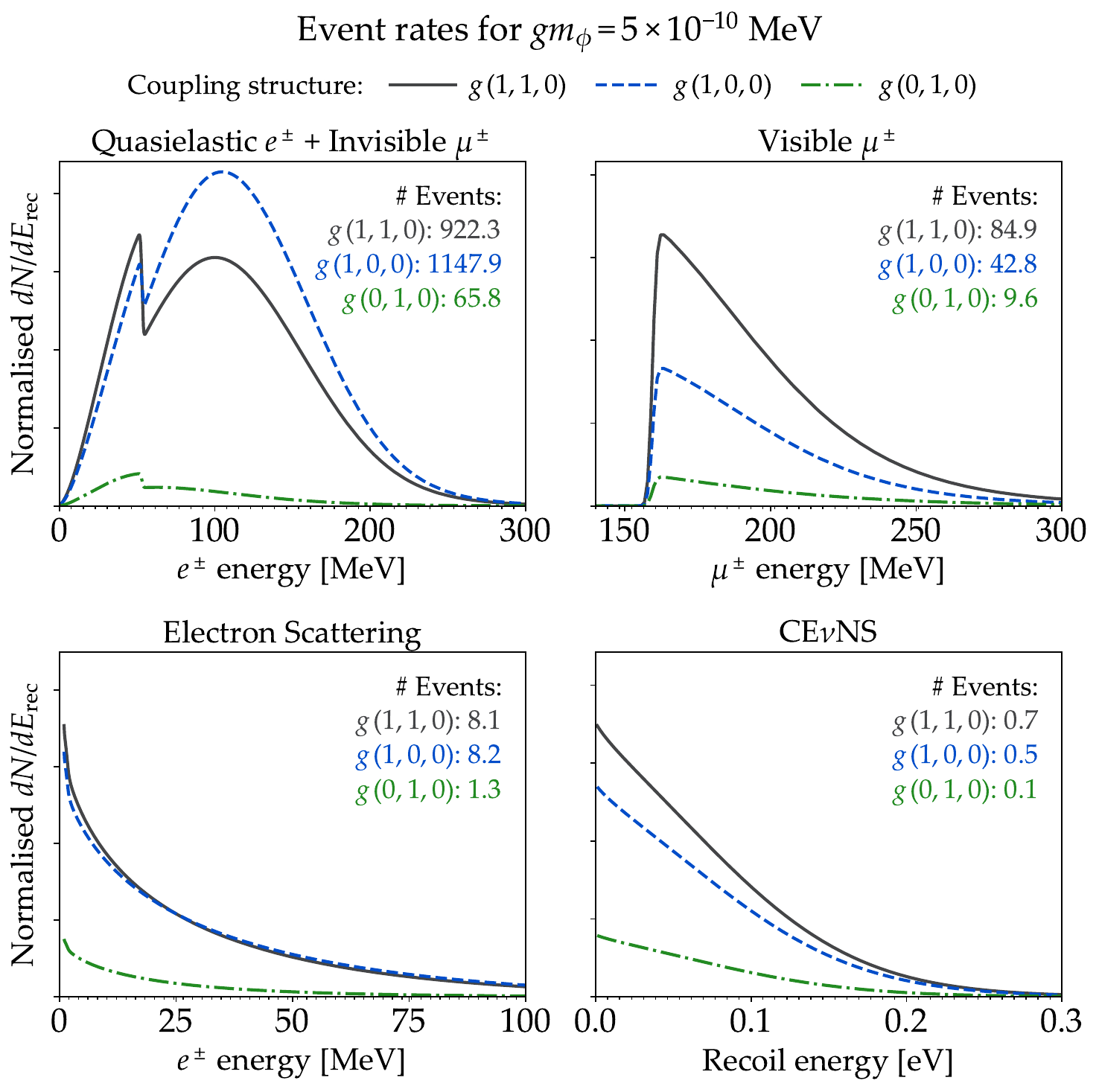}
    \caption{\textbf{\textit{Flavor-dependent analysis benchmark event rates in each of the detection topologies.}} We simulate a light Majoron with a mass of $m_\phi = 1$~MeV, with couplings allowed by current SN~1987A bounds of $g m_\phi = 5\times 10^{-10}$ MeV. For the pessimistic choice $g m_\phi=2\times 10^{-10}$ MeV (not shown), the flux trivially scales with the square of the coupling. Events were simulated assuming a cold SN model.
    }
    \label{fig:flavor_coupling_reconstruction_benchmark}
\end{figure*}

Figure~\ref{fig:flavor_coupling_reconstruction_benchmark} shows the event rates for all three cases. First, there is a clear dependence of the total number of events on the coupling structure, which is understood based on our earlier analysis of the Majoron signal: for the cold model, a nonzero coupling to $\nu_e$ only leads to a much larger event rate than a nonzero coupling to $\nu_\mu$. When only $g_e \neq 0$, the event rates are larger than when both $g_e$ and $g_\mu$ are nonzero and equal, since in the latter case, for the cold model, the additional small Majoron emission from $\nu_\mu \nu_\mu$ coalescence in the SN is offset by the enhanced fraction of $\overline{\nu}_e$ for Majoron decay to  $\overline{\nu}_e + \overline{\nu}_e$. 

However, the absolute number of events cannot discriminate different coupling structures, since it is degenerate with the unknown absolute value of the coupling. Luckily, we easily identify other features of the signal that break the degeneracy: the height of the peak of Michel electrons and positrons relative to the peak of the event rate in the QE $e^\pm$ + Inv.~$\mu^\pm$ channel, and  number of events in the visible $\mu^\pm$ channel are good proxies of the fraction of muon neutrinos at Earth. 
The number of events in the ES channel could also be a good proxy of the all-flavor flux, but the event rates are low even for our optimistic benchmark case with the largest coupling. Finally, for CE$\nu$NS, less than one event is expected in all cases, so this channel does not play a role in our analysis.

For the statistical analysis, we use the same definition of the likelihood function as before, in Eq.~\eqref{eq:sim_likelihood}, except that it now depends on $g_e$ and $g_\mu$ separately, and the true flux now includes the standard SN neutrinos from the cold SN model plus the neutrinos from the decay of a light Majoron with $m_\phi=1$~MeV (like before, we do not profile the likelihood over $m_\phi$).  We consider two cases: an optimistic one with $g m_\phi = 5 \times 10^{-10}$~MeV and a pessimistic one with $g m_\phi = 2 \times 10^{-10}$~MeV. For a fixed value of $g m_\phi$, the outcome is essentially independent of $m_\phi$ unless it is of tens of MeV, where Majoron production starts to be kinematically suppressed. Hence, our projected reconstruction is valid for any Majoron mass lower than tens of MeV and larger than tens of keV, below which it becomes comparable to the neutrino thermal mass; see Section~\ref{sec:neutrinos_from_Majorons}. 
The test statistic, TS, is given by Eq.~\eqref{eq:definition_ts}, evaluated now at test values of $g_e$ and $g_\mu$ separately. Since we are no longer setting upper bounds on the couplings (Section~\ref{sec:projected_bounds}), but measuring their values, the TS is expected to follow a $\chi^2$ distribution with two degrees of freedom, $g_e m_\phi$ and $g_\mu m_\phi$, and so the threshold value for $95\%$~C.L. is TS = 6.

\begin{figure*}
    \centering
    \includegraphics[width=0.95\textwidth]{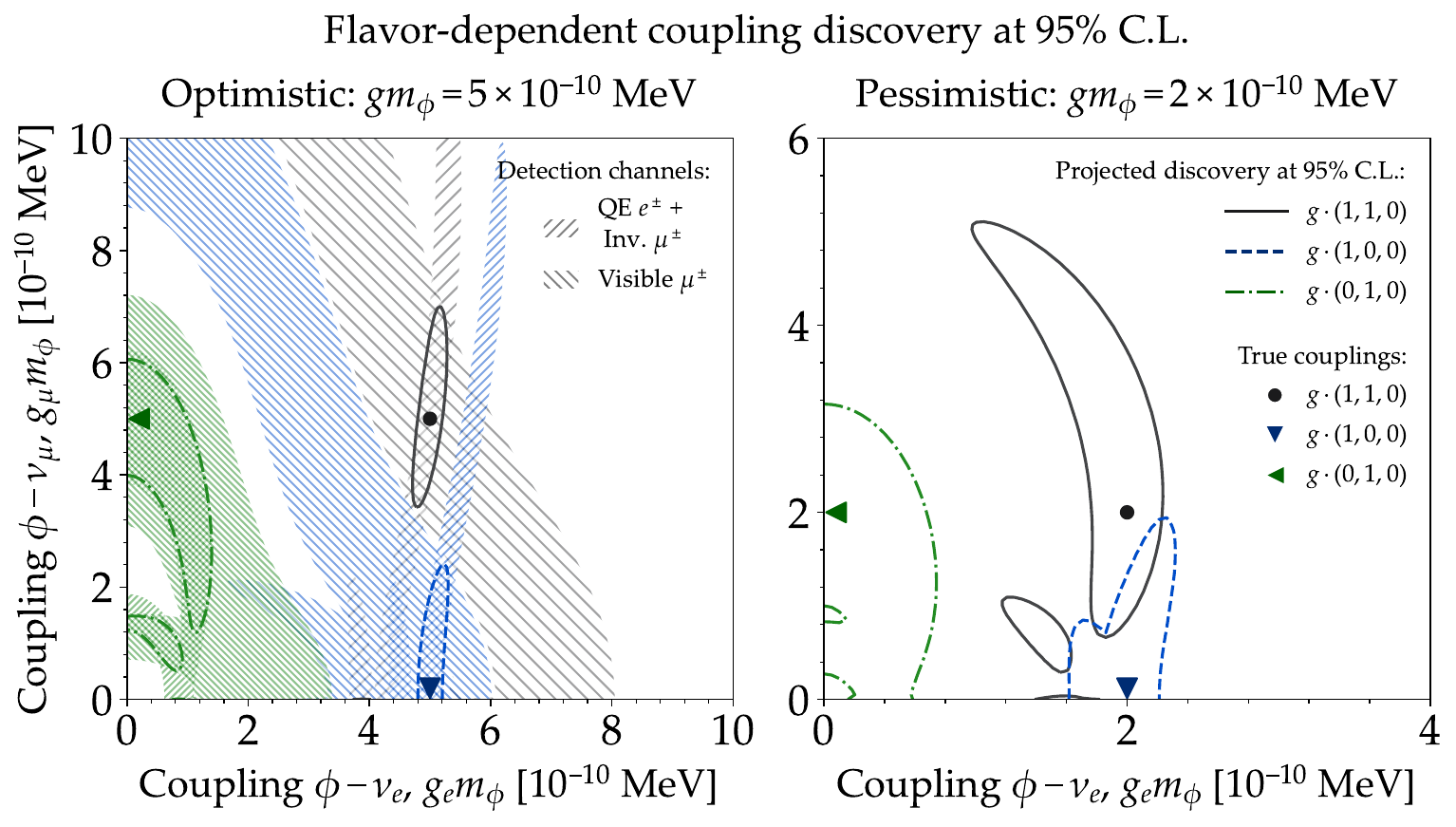}
    \caption{\textbf{\textit{Discovery of a Majoron-neutrino flavor-dependent coupling.}} We show the $95\%$~C.L.~discovery regions for different flavor textures of the Majoron-neutrino coupling, assuming optimistic (\textit{left}) and pessimistic (\textit{right}) illustrative values of the coupling.}
    \label{fig:flavor_discovery}
\end{figure*}

Figure~\ref{fig:flavor_discovery} shows the resulting projected joint allowed regions of $g_e m_\phi$ and $g_\mu m_\phi$, for the above optimistic and pessimistic benchmarks of their true values, and assuming three sample flavor textures: pure $g_e$ coupling, pure $g_\mu$ coupling, and $g_e = g_\mu$ couplings.
For the optimistic benchmark, we also show separately the discovery contours using only events in the QE $e^\pm$ + Inv.~$\mu^\pm$ channel and in the visible $\mu^\pm$ channel, to highlight their contributions. 
For this benchmark, the large number of detected events allows for a precise measurement of the couplings, thanks to the complementary information carried by the different topologies. The QE $e^\pm$ + Inv.~$\mu^\pm$ channel yields a narrow interval of $g_e$, since even a tiny increase or decrease in $g_e$ would dramatically alter the expected number of high-energy neutrino events in this topology. However, this channel alone provides nearly no sensitivity to $g_\mu$, which is provided by visible $\mu^\pm$.  Moreover, \figu{flavor_discovery} shows that the three sample flavor textures can be discriminated from one another.

For the pessimistic benchmark, the event rates are reduced by more than a factor~$4$ due to the lower coupling, and the allowed regions of $g_e m_\phi$ and $g_\mu m_\phi$ are widened. The allowed regions are characteristically multi-humped, and in the $g (1,1,0)$ they become even disconnected, due to profiling the likelihood function over the cold and hot SN models. As we stated out earlier, even though we adopt the cold SN model as the true one when generating our mock observations, we assume no knowledge of the SN model in the statistical analysis.  This allows us to explain the mock observations also using the hot SN model with lower values of the couplings compared to those preferred by the cold SN model. As pointed out in Section~\ref{sec:projected_bounds}, this is an overly conservative step, since in reality the information from the low-energy standard SN events would allow us to resolve the mass of the PNS, and to discriminate between the hot and cold models. Regardless, even in the pessimistic benchmark, the three sample flavor textures may be discriminated from each other. 

\subsection{Time-dependent, flavor-insensitive analysis}
\label{sec:time_dependent_flavor_insensitive}

\begin{figure*}[t!]
    \centering
    \includegraphics[width=0.8\textwidth]{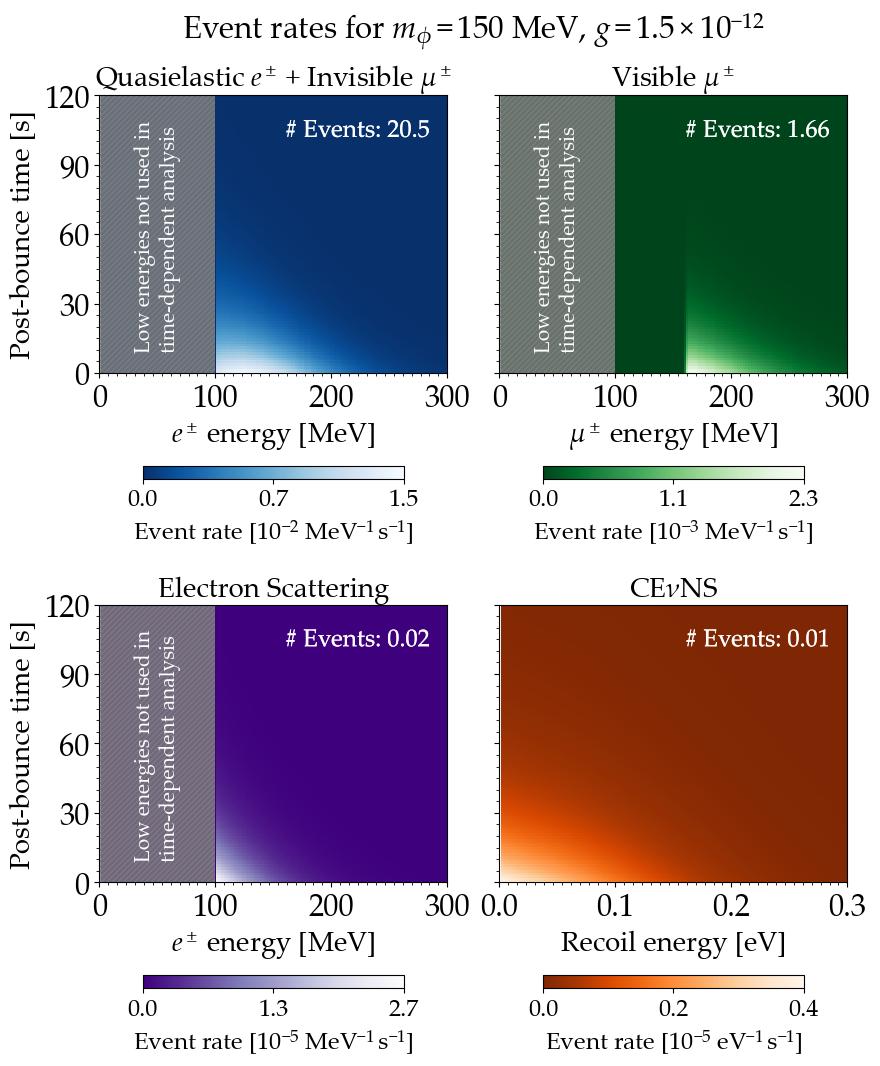}
    \caption{\textit{\textbf{Timing-dependent analysis benchmark event rates for each detection topology.}} We simulate a Majoron of $m_\phi = 150$ MeV with flavor-universal coupling $g = 1.5\times 10^{-12}$. The gray bars indicate the energy cut of our analysis that removes events below $100$ MeV in order to eliminate the standard SN neutrino flux. We show the electron scattering and CE$\nu$NS event rates here, but do not include them in our analysis.
    }
    \label{fig:mass_coupling_reconstruction_benchmark}
\end{figure*}

Majorons are produced mostly in the very first second post-bounce in the SN core, when neutrinos are strongly degenerate. Therefore, if a high-energy neutrino flux from their decay were to be observed over a much longer duration, at several tens of seconds post-bounce, this would be a natural indication that the Majoron is heavy. Measuring the time scale would allow us to break the degeneracy between the Majoron coupling and mass that would otherwise exist. 
Crucially, however, detecting neutrinos at these late times is \textit{not} by itself a unique signature of the presence of Majorons, since standard late-time SN emission from fallback accretion is possible (Section~\ref{sec:neutrino_detection-experiments_other}). Only by virtue of their high energies we can be sure of their origin. 

Disentangling the neutrinos from Majoron decay from the standard SN neutrinos is a more complex task in a time-dependent analysis than in our earlier time-integrated one. While the standard neutrino fluence in the cooling phase is described well by Eq.~\eqref{eq:nufromSN}, there is no available parametrization of the time-dependent flux. This situation is even worse for the late-time neutrino emission. We circumvent this limitation by 
cutting from our analysis all events with energies below $100$~MeV, so that the expected number of events from Majoron decay across all topologies is $52.4$ for the hot model and 22.2 for the cold model, while the expected number of standard SN events is, respectively, only $6.5$ and 0.05, less than the Poisson fluctuations on our signal. The potential contamination from late-time fallback accretion has typically even lower energy than during the cooling phase. Thus, in our analysis we neglect the standard SN contamination above 100~MeV.
This hard cut in energy is conservative, meant to show the potential of a time-dependent analysis. In reality, below $100$~MeV the neutrinos from Majoron decay exhibit unique signatures, such as the characteristic shape of the spectrum of Michel positrons from the decay of higher-energy muons, that could be used to identify them even in the absence of a parametrization of the time-dependent standard SN flux.

Figure~\ref{fig:mass_coupling_reconstruction_benchmark} shows the distributions in time and energy of the expected events, for the cold SN model, and for benchmark choices of the model parameters.  We adopt a benchmark heavy Majoron with flavor-universal couplings of $g_e = g_\mu = g_\tau \equiv g = 1.5 \times 10^{-12}$ and mass of $m_\phi = 150$~MeV. 
The only two topologies with a detectable number of events are QE $e^\pm$ + Inv.~$\mu^\pm$ and visible $\mu^\pm$. As the coupling considered is small, electron scattering and CE$\nu$NS yield negligible event rates. 

Since we discard the standard neutrinos above 100~MeV, only the neutrinos from Majoron decay contribute to the event rates. We define the likelihood function in analogy to Eq.~\eqref{eq:sim_likelihood}, but now using the joint energy and time distribution of the events, i.e.,
\begin{equation}\label{eq:sim_likelihood_time_dep}
\begin{split}
    \chi^2(g,m_\phi)
    =
    & -2
    \, \mathrm{max}_{\boldsymbol{\theta}_\mathrm{SN}}\langle \log\mathcal{L}\left(g,m_\phi; \boldsymbol{\theta}_{\rm SN}\right) \rangle
    \\ 
    =   
    & -2 \sum_{\rm t}
    \left[
    \int dt \int_{90\;\mathrm{MeV}}^{E_{\mathrm{max}}}dE_{\rm rec}\, \frac{dN_{\rm t,true}}{dt dE_{\rm rec}}\, \text{ln}\left(\frac{dN_{\rm t,test}(g,m_\phi;\boldsymbol{\theta}_\mathrm{SN})}{dtdE_{\rm rec}}\right) 
    \right. 
    \\
    & \left. - \int \, dt \int_{90\;\mathrm{MeV}}^{E_{\mathrm{max}}}dE_{\rm rec}\, \frac{dN_{\rm t, test}(g,m_\phi;\boldsymbol{\theta}_\mathrm{SN})}{dtdE_{\rm rec}}
    \right] \;,
\end{split}
\end{equation}
where now $\boldsymbol{\theta}_{\rm SN}$ represents the choice between the cold and hot SN models. In analogy to Eq.~\eqref{eq:definition_ts}, the test statistic is
\begin{equation}
    \label{eq:ts_time_dep}
    \mathrm{TS}(g,m_\phi)=\chi^2(g,m_\phi)-\chi^2(\hat{g},\hat{m}_\phi) \;,
\end{equation}
where, unlike our earlier analyses, we now minimize TS simultaneously over $g$ and $m_\phi$; in Eq.~\eqref{eq:ts_time_dep}, $\hat{g}$ and $\hat{m}_\phi$ are their best-fit values, which coincide with the true values. We evaluate the true event spectrum at the same benchmark choices of the model parameters as in \figu{mass_coupling_reconstruction_benchmark} and the test event rate at test values of $g$ and $m_\phi$. The TS is obtained by profiling over the SN hot and cold models; it is expected to follow a $\chi^2$ distribution with two degrees of freedom, so the regions of $g$ and $m_\phi$ allowed at $95\%$~C.L.~are obtained by setting $\mathrm{TS} = 6$. 

\begin{figure}
    \centering
    \includegraphics[width=0.60\textwidth]{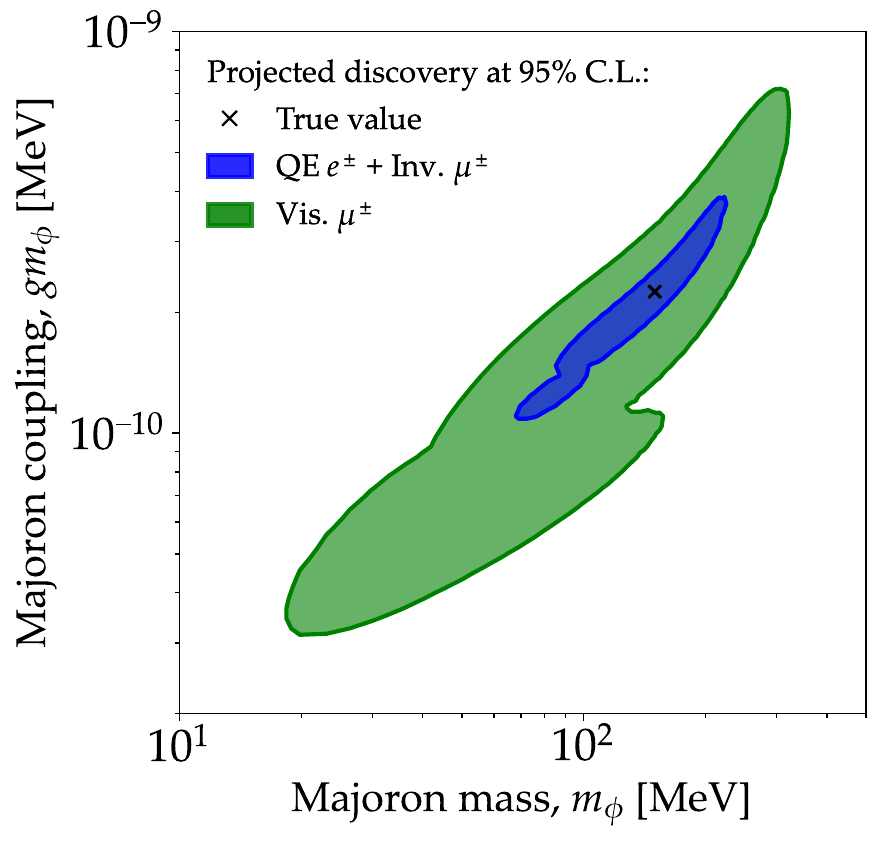}
    \caption{\textit{\textbf{Majoron mass reconstruction.}}  The measurement is based on our time-dependent, flavor-insensitive analysis strategy.  The true, simulated Majoron mass and flavor-universal coupling are $m_\phi = 150$ MeV and $g = 1.5\times 10^{-12}$.  We show separately the result of using the two topologies with the highest event rates:  QE $e^\pm$+Inv. $\mu^\pm$ and visible $\mu^\pm$. 
    }
    \label{fig:timing_discovery}
\end{figure}

Figure~\ref{fig:timing_discovery} shows the $95\%$~C.L.~allowed regions of the Majoron mass and coupling. We show separate contours for the two visible detection channels, but the sensitivity of reconstruction is driven primarily by the QE $e^\pm$ + Inv.~$\mu^\pm$ channel due to its much higher rate. The Majoron mass is reconstructed to within about 60\% of its true value. The allowed region is two-humped, with the two humps resulting from profiling over the hot and cold SN models.  This is because, even if the simulated SN assumed the cold model, the observed SN signal can also be explained with a lower coupling if the original SN model was hot. As stated previously, in reality, thanks to the copious low-energy SN neutrinos, we should be able to assert whether the model was truly cold or hot, and thus the allowed region would be smaller.

\section{Summary and outlook}

It is general agreement that a copious detection of neutrinos from the next galactic supernova (SN) will be a powerful testbed for new physics. However, this statement is sometimes used as an overly optimistic motto, especially since many of the observables currently used to probe new physics would not benefit significantly from a signal seen with a large number of detected events. A paradigmatic example is the celebrated SN~1987A cooling bound~\cite{Raffelt:1996wa}, based on the duration of the neutrino burst. This constraint is presumably not limited by experimental uncertainties---after all, the detected neutrino events were sufficient to pinpoint the duration of the SN~1987A burst certainly to within a factor~of~2---but rather by theoretical ones, since convection and neutrino-nucleon opacities seem to play the key role in shaping the duration of the cooling phase~\cite{Fiorillo:2023frv}, an aspect unaccounted for in the historical studies of the cooling bounds. The main possibility to learn about beyond the Standard Model (BSM) physics from the boon of neutrino data expected from the next galactic SN is to directly detect the neutrinos produced in the decay of Majorons or other feebly interacting particles. This strategy would benefit from a large event rates, leading either to much more stringent bounds or to a discovery.
Making best use of this possibility requires a concrete strategy relating the properties of the BSM model to the observables of the neutrino signal, marginalizing over our ignorance of the SN physics details. This is the open question we tackled in this work. 

We have considered the production of high-energy Majorons via the coalescence of neutrinos inside the proto-neutron star (PNS) of a SN. For this setup, the lepton number violation maximizes the energy of the BSM neutrinos, which are produced primarily from the coalescence of $\nu_e$ with large chemical potential. The subsequent decay of the Majorons into neutrinos leads to a high-energy component in the flux observable at Earth. Each neutrino event detected in ongoing and planned neutrino and dark matter experiments will be characterized by a limited number of features, namely its energy, the event topology in the detector, and the time of arrival. Compared to the existing literature, we exploit for the first time all these features to reconstruct the properties of the underlying BSM model, \ie, the mass of the Majoron and its couplings to $\nu_e$, $\nu_\mu$, and $\nu_\tau$.

We have geared our forecasts to the detection of SN neutrinos by upcoming observatories Hyper-Kamiokande and DARWIN.  They provide sensitivity to multiple neutrino detection channels: Hyper-Kamiokande, to quasi-elastic (QE) $\nu_e$ and $\nu_\mu$ scattering, and to neutrino--electron ES, and DARWIN, to coherent elastic neutrino-nucleon scattering (CE$\nu$NS).  Combined, the different detection channels grant us partial sensitivity to the flavor composition of the SN neutrino flux.  Concretely, we gain access to the content of $\nu_e$ and $\nu_\mu$ in the flux by contrasting the number of detected high-energy visible muons, the number of events from $\nu_e$ QE scattering, and the height of the peak of Michel $e^\pm$ coming from invisible muon decays relative to the remaining $\nu_e$ QE scattering events. At high energies, the expected rate of CE$\nu$NS events is, unfortunately, negligible.

The only observable that allows us to unambiguously distinguish the standard SN neutrino flux from the neutrinos produced by Majoron decay is the energy spectrum. As pointed out in Refs.~\cite{Akita:2022etk, Fiorillo:2022cdq}, neutrinos from Majoron decay have energies higher ($\sim 100\,\rm MeV$) than standard SN neutrinos ($\sim 10\,\rm MeV$). The possibility of measuring the energy of individual neutrinos allows us to obtain results that depend only weakly on the uncertainties in the standard SN neutrino production.

We have shown the potential use of the signal from the next galactic SN via three separate kinds of analyses. In the first analysis, we focus on the negative viewpoint that no high-energy neutrino will be observed, in which case we will only obtain bounds on BSM physics. While this is a possibility, the best-case scenario would of course be the one in which a signal is observed. Therefore, our second and third analyses refer to the case in which a high-energy neutrino signal will be detected from the next galactic SN. Below, we detail each of these analyses.

First, in case no event with a particularly high energy is observed, we can constrain Majoron couplings to neutrinos of different flavors. Such bounds rely on two observables---the energy spectrum and the topologies of the detected events. We have revisited the projected bounds on heavy Majorons by considering two largely different cases for the PNS profiles of density and temperature, a cold and a hot simulated model. The bounds we find on the couplings are consistent with simple order-of-magnitude estimates, and highlight the dependence on the flavor texture of the couplings. In our analysis we are overly conservative, assuming that we have no knowledge of the mass of the PNS, which determines the typical temperature and luminosity of the model; practically, we always profile over the hot or cold model.
In reality, the large standard neutrino signal in the tens-of-MeV energy range would allow us to pinpoint the mass of the PNS and potentially improve the bounds, especially if the next galactic SN involves a heavy PNS.

Second, for a large coupling, many neutrinos from Majoron decays will be observed in all the different topologies within a span of a few seconds. For this scenario, we show that the combining different event topologies allows us to reconstruct the separate couplings $g_e m_\phi$ and $g_\mu m_\phi$. Extending this analysis to more general coupling structures would presumably be more useful to pursue in case of a real observation; here we show as a proof of principle that a partial flavor structure reconstruction of the model is possible with the expected event rates, focusing on couplings of the order of $g_\alpha m_\phi \sim 10^{-10}$~MeV.

Third, for large masses and low couplings (our benchmark is a mass $m_\phi=150\,\rm MeV$ and a flavor-universal coupling $g=1.5\times 10^{-12}$), the time distribution of the detected events can be stretched by up to tens of seconds. Folding both the time and energy distributions in our analysis allows us to jointly measure the Majoron mass and coupling. The current absence of a parametrization of the time dependence of the standard SN neutrino production motivates us to discard events with energies below 100~MeV from our analysis in order to remove the otherwise dominant standard signal. We show that the timing information can be used to infer the BSM parameters, while the high energy of the events involved guarantees minimal-to-no sensitivity to the SN physics uncertainties.

As a general comment, we are of course unaware of what properties the next galactic SN will have. However, most of our results turn out to be qualitatively independent of this ignorance. For all of our reconstruction prospects, we consider that the next galactic SN will resemble our cold model, which is a pessimistic choice, since a hotter SN would produce a larger signal and allow us to reconstruct the Majoron properties more precisely. We assume that the SN will occur 10~kpc away; for the projected bounds that we draw, the bounds on the couplings scale linearly with the distance. Similarly, if a SN should explode before Hyper-Kamiokande starts operations, one can still benefit from the exposure of its predecessor Super-Kamiokande~\cite{Super-Kamiokande:2002weg}, which is currently active. In that case, the bounds on $g_\alpha$ would be worsened by the square root of the effective volume ratio for the two experiments. For the projected reconstruction of the BSM parameters, the impact of a different distance must be assessed by a dedicated analysis.  Finally, the next galactic SN might fail to produce a neutron star, with the remnant quickly collapsing into a black hole and interrupting neutrino emission during the cooling of the PNS. This would be to our advantage, since the Majorons would still be produced within the first second, before the collapse to a black hole, while the standard neutrino emission would be cut off, reducing the background for BSM searches.

Overall, for the first time, we have shown that the multiple topologies and timing of the events detected from the next galactic SN could allow us to reconstruct the mass and couplings of Majorons that decay into neutrinos. Our findings can be applied to a plethora of different models positing particles that decay into neutrinos, such as novel gauge bosons or sterile neutrinos.

\section*{Acknowledgments}

We thank Thomas Janka and Georg Raffelt for useful comments, and Kensuke Akita, Jeffrey Lazar, Ying-Ying Li, Carlos Arg\"{u}elles, and Vedran Brdar for friendly discussions. MB, DFGF, and BT are supported by
Villum Fonden under project no.~29388. DFGF acknowledges support from the European Union’s Horizon 2020 Research and Innovation Program under the Marie Sklodowska-Curie Grant
Agreement No.~847523 ``INTERACTIONS.'' PMM acknowledges support from the Carlsberg Foundation (CF18-0183). EV is supported by the Italian MUR Departments of Excellence grant 2023-2027 ``Quantum Frontiers'' and by Istituto Nazionale di Fisica Nucleare (INFN) through the Theoretical Astroparticle Physics (TAsP) project. This work used the Tycho supercomputer hosted at the SCIENCE High Performance Computing Center at the University of Copenhagen. 

\appendix

\section{Time distribution of decay neutrinos}\label{app:time_distribution}

In this appendix we derive explicitly the time distribution for the neutrinos produced in the decay of a Majoron. The geometry of the process is illustrated in Fig.~\ref{fig:drawing}. We assume the Majoron has an energy $E_\phi$ and thus moves with a velocity $v_\phi=\sqrt{1-m_\phi^2/E_\phi^2}$; we do not assume ultra-relativistic Majorons.

The time delay of a neutrino from the decay of a Majoron, compared to the arrival time of a neutrino moving in straight line, is
\begin{equation}\label{eq:time_delay_1}
    t=\frac{\ell}{v_\phi}+s-D_\mathrm{SN} \;,
\end{equation}
where $\ell$ is the distance that the Majoron travels before it decays and $s$ is the distance that its daughter neutrinos travel from there to the Earth.
We can safely assume that $\ell \ll D_\mathrm{SN} \sim~{\rm kpc}$, since $\ell$ is roughly the size of the decay length of the boosted Majoron, i.e., $\ell_\phi \sim E_\phi/\sum_\alpha g_{\alpha}^2 m_\phi^2$. Even for $g m_\phi\ll 10^{-12}$~MeV, much lower than what we will be able to probe with future experiments, for a Majoron energy of $E_\phi\sim 100$~MeV, this yields $\ell_\phi \simeq 2\times 10^{15}~{\rm cm} \ll D_\mathrm{SN}$. Thus, we can approximate $s\simeq D_{\rm SN}-\ell \cos\alpha \gg \ell$, where $\alpha$ is the angle at which the Majoron is emitted from the PNS, measured with respect to the straight-line distance from the PNS to Earth. It also follows from here that the angle with which the neutrinos arrive at Earth, also measured with respect to the straight-line distance from the PNS to Earth is $\beta \ll 1$; concretely,
\begin{equation}
    \beta\simeq\frac{\ell\sin\alpha}{s}\simeq \frac{\ell\sin\alpha}{D_{\rm SN}} \;.
\end{equation}
With this approximation, the angle $\alpha$ of emission of the Majoron and the angle $\theta$ between its direction and the direction of the neutrinos are nearly identical, since $\beta \ll \alpha$, i.e., $\theta \simeq \alpha$.

We now connect the angle of emission with the energy of the emitted neutrinos. From the kinematics of the Majoron decay, we find that
\begin{equation}
    \cos\theta=\frac{1}{v_\phi}\left(1-\frac{m_\phi^2}{2E_\phi E_\nu}\right) \;.
\end{equation}
Replacing $s$ and $\alpha$ in Eq.~\eqref{eq:time_delay_1} yields
\begin{equation}
    \label{eq:delay_time_def}
    t=\frac{\ell m_\phi^2}{2E_\nu E_\phi v_\phi} \;.
\end{equation}
The time delay is proportional to the length, $\ell$, travelled by the Majoron before it decays. As is well known, this length has a probability distribution $\propto e^{-\ell/\ell_\phi}$, i.e., 
\begin{equation}
    P(\ell)d\ell=\exp\left(-\frac{\Gamma_\phi m_\phi \ell}{E_\phi v_\phi}\right)\frac{\Gamma_\phi m_\phi}{E_\phi v_\phi}d\ell \;,
\end{equation}
where we have multiplied the decay rate, $\Gamma_\phi = 1/\ell_\phi$, by the Lorentz boost factor $E_\phi/m_\phi$. From here, using Eq.~\eqref{eq:delay_time_def}, the distribution of time delays is
\begin{equation}
    P(t) dt=\exp\left[-\frac{2\Gamma_\phi E_\nu t}{m_\phi}\right]\frac{2\Gamma_\phi E_\nu}{m_\phi} dt \;.
\end{equation}
Since this distribution does not depend on the Majoron energy, but only on the neutrino energy, we can simply multiply it by the neutrino energy spectrum to obtain the joint time and energy distribution of the neutrinos that reach Earth, recovering the result in Eq.~\eqref{eq:ndot-nu} in the main text.

\bibliographystyle{bibi}
\bibliography{References}

\end{document}